\documentclass[letterpaper,11pt]{article}
\usepackage[left=2.5cm,right=2.5cm,top=2.75cm,bottom=2.5cm,nohead]{geometry}

\usepackage{amsmath,amssymb,color,graphicx,url}

\usepackage{mathptmx}
\usepackage{courier}
\usepackage[scaled=.90]{helvet}

\def\emdash{---}
\def\etal{{\em{et al}}}

\def\Celegans{{\it{C.\ elegans}}}

\def\hmtimes{\times}
\def\hmsum{\sum}

\def\hmrarr{\rightarrow}

\def\hmforall{\forall}

\def\hmlang{\langle}
\def\hmrang{\rangle}
\def\hmleq{\leq}

\def\hmneq{\neq}

\def\hmisin{\in}

\def\hmBeta{\mbox{\rm{B}}}
\def\hmbeta{\beta}
\def\hmChi{\mbox{\rm{X}}}
\def\hmchi{\chi}
\def\hmDelta{\Delta}
\def\hmdelta{\delta}

\def\hmGamma{\Gamma}
\def\hmgamma{\gamma}

\def\hmLambda{\Lambda}
\def\hmlambda{\lambda}
\def\hmMu{\mbox{\rm{M}}}
\def\hmmu{\mu}

\def\hmmicron{\mu{}m}

\def\hmPi{\Pi}
\def\hmpi{\pi}
\def\hmPhi{\Phi}
\def\hmphi{\phi}
\def\hmPsi{\Psi}

\def\urlh#1#2{{\footnote{\url{#1}}}}

\def\figures{./}

\def\redhighlight#1{\textcolor{red}{#1}}

\def\rasterisk{\textcolor{red}{*}}

\newcommand{\R}{\mathbb{R}}        
\newcommand{\F}{\mathcal{F}}       
\newcommand{\B}{\mathcal{B}}       
\newcommand{\C}{\mathcal{C}}       
\newcommand{\Domains}{\mathcal{D}} 
\newcommand{\Ranges}{\mathcal{R}}  

\def\hmFs{\hmPsi{}} 

\def\hmps{\hmphi{}} 
\def\hmPs{\hmPhi{}} 

\def\hmgp{{\hmmu{}}} 
\def\hmGp{{\hmMu{}}} 

\def\hmbf{\hmbeta{}}
\def\hmBf{\hmBeta{}}

\def\hmcp{{\hmpi{}}}
\def\hmCp{{\hmPi{}}}

\def\hmgbp{\hmdelta{}}
\def\hmGbp{\hmDelta{}} 

\def\hmdgf{\hmgamma{}}
\def\hmDgf{\hmGamma{}}

\def\hmdsf{\hmlambda{}} 
\def\hmDsf{\hmLambda{}} 

\def\hmdcf{\hmchi{}} 
\def\hmDcf{\hmChi{}} 

\title{Inferring Mesoscale Models of Neural Computation}

\date{}

\author{{\bf{Thomas Dean}}\\Google Research}

\begin{document}




\pagenumbering{gobble}

\begin{titlepage}

  \maketitle

  \begin{abstract}
    Recent years have seen dramatic progress in the development of techniques for measuring the activity and connectivity of large populations of neurons in the brain.  However, as these techniques grow ever more powerful{\emdash{}}allowing us to even contemplate measuring every neuron in entire brains{\emdash{}}a new problem arises: how do we make sense of the mountains of data that these techniques produce? Here, we argue that the time is ripe for building an intermediate or {\em{mesoscale}} computational theory that can bridge between single-cell (microscale) accounts of neural function and behavioral (macroscale) accounts of animal cognition and environmental complexity. Just as digital accounts of computation in conventional computers abstract away the non-essential dynamics of the analog circuits that implementing gates and registers, so too a computational account of animal cognition can afford to abstract from the non-essential dynamics of neurons. We argue that the geometry of neural circuits is essential in explaining the computational limitations and technological innovations inherent in biological information processing. We propose a blueprint for how to employ  tools from modern machine learning to automatically infer a satisfying mesoscale account of neural computation that combines functional and structural data, with an emphasis on learning and exploiting regularity and repeating motifs in neuronal circuits. Rather than suggest a specific theory, we present a new class of scientific instruments that can enable neuroscientists to propose, implement  and test mesoscale theories of neural computation. 
  \end{abstract}

\end{titlepage}

\newpage

\pagenumbering{roman}

\tableofcontents

\newpage

\pagenumbering{arabic}





\label{sec_introductory_remarks}
\section{Introduction}

In the past decade, experimental neuroscience has been transformed by a veritable ``perfect storm'' of technological advances, ranging from powerful molecular and optical tools that allow us to record and manipulate ever larger populations of neurons with single-cell resolution~\cite{NobaueretalNATURE-METHODS-17,NguyenetalBIORXIV-17,FriedrichetalPLoS-17,PrevedeletalNATURE-METHODS-16,SadakaneetalCELL-15}, to technologies that allow us to reconstruct the wiring of neuronal circuits down to the level of synapses~\cite{HildebrandetalNATURE-17,LeeetalNATURE-16,KasthurietalCELL-15,HayworthetalNATURE-METHODS-15,EberleetalJM-15,BocketalNATURE-11}. While systems neuroscience experiments were once limited to recording one or a few neurons at a time, we can now even begin contemplating recording from {\it{every}} neuron in an organism~\cite{DupreandYusteCURRENT-BIOLOGY-17,FriedrichetalBIORXIV-16,NguyenetalPNAS-15,AimonetalBIORXIV-15,AhrensetalNATURE-13}. In many ways, neuroscience has entered a new era in which we are increasingly no longer limited by the data we can collect, but rather are limited by our ability to process, understand, and reason about the data we collect. Science fundamentally progresses through the construction of models: models that explain, simplify and abstract away detail from experimental data. As our reach expands towards ever more comprehensive collected data, we must develop new tools for understanding that data. 

In this paper, we argue that it is time to begin tackling the problem of {\it{mesoscale}} models of neuronal systems, with the goal of building frameworks that can accommodate up to whole organism-level neuronal models. We propose a strategy for thinking about biological modeling and hypothesis testing, a class of functional mesoscale models for whole organisms, and a road map for building the required technology and infrastructure. We focus primarily on describing the proposed class of mesoscale models and the machine-learning architecture necessary for inferring such models. Our proposed models are {\it{compositional}} models in which the components are artificial neural networks, characterized here as {\it{shallow deep-neural-networks}} for reasons that will soon become apparent.




\label{subsec_mesoscale_models}
\subsection{Mesoscale Models}

What is a \emph{mesoscale} model? Put simply, a mesoscale model (or mesoscale theory) is an explanatory framework that exists somewhere between the micro and macro scales. In the context of neuroscience, a mesoscale theory is intended to serve as a conceptual and explanatory bridge between cellular and molecular theories on the one hand, and cognitive and behavioral theories on the other. A micro-scale model of a neuron might be satisfying if it explains the internal molecular and electrical function of the cell, without regard to what role it plays in a larger computation supporting behavior. Likewise, a cognitive model can help explain a behavior without necessarily explaining how populations neurons give rise to the computations underlying that behavior. The purpose of a mesoscale model is to explicitly bridge between these scales \emdash{} explaining how the activity and structure of a neuronal network give rise to useful computation and behavior.

The concept of mesoscale modeling is not unique to neuroscience; many fields of science and engineering explicitly or implicitly invoke mesoscale models. In computer science, programmers rarely concern themselves with the semiconductor physics of silicon, or the behavior or transistors, or even the construction of logic gates and larger structures such as half-adders. Instead, computer languages sit atop a nested pyramid of abstractions, each layer exposing simpler abstractions atop the incredible multi-scale complexity that lies beneath. Abstractions such as ``stack,'' ``heap,'' and ``cache'' provide the programmer with essential intuition about how the computer will run their code, without requiring them to understand details of their hardware implementation. Similarly, in fluid dynamics, the Navier-Stokes equations describe the movement of fluids, abstracting away from the behavior of individual molecules in the fluid, but stops short of trying to directly encapsulate higher level concepts, such as aero- or hydrodynamic lift. Each level of abstraction in these example model hierarchies have their own different mathematics, terminology and physical intuitions. A good mesoscale theory describes {\em{emergent}} behavior, allowing for the aggregate behavior of elements at lower levels to be collapsed into coarser-grain approximation that explain the behavior aggregate in simpler terms. A mesoscale theory that can provide a compelling bridge between the extremes of scale evident in neural function is arguably a holy grail in computational neuroscience.

What are the requirements for building such a mesoscale model in neuroscience?  We argue that a key feature of a mesoscale neuronal model is that it should be able to take in functional activity measurements \emdash{} ideally starting from as close to the sensory periphery as possible \emdash{} along with anatomical information data about the connectivity among neurons (``connectomes'') and produce predictions about the activity of other neurons in the system \emdash{} ideally including neurons close to the motor outputs of the system. The goal of such a model is to provide as close to an end-to-end description of the system's input-output transfer function, while respecting (and offering explanation for) the internal organization of the system at an intermediate scale.

We note that this approach is different from other kinds of models employed in neuroscience and psychology today.  For instance, psychophysical models seek to describe the input-output transfer function of an organism without necessarily worrying about how computations are specifically implemented in the brain. Traditional theoretical neuroscience models typically build from the ground up using a collection of simpler principles (e.g. excitation-inhibition balance, all-to-all connectivity, etc.), and then endeavor to explain phenomena observed in data. Single unit biophysical models focus primarily on the detailed working of individual units, and larger-scale biophysical models simulate the consequences of simulating larger populations connected together according to simulated connectivity.

Our proposed mesoscale models, which are described below in detail, embrace intermediate-level abstraction, encapsulating the complexity of neuronal subcircuits using modern machine learning tools, but also producing an end-to-end system that can be ``run'' with new inputs, yielding new predicted outputs and internal states. Building such models will not be easy, but we believe that for the first time, data and computational tools exist (or nearly exist) that could enable the attempt. We also note that we have no illusions that such models will be ``correct'' at the outset \emdash{} indeed, one of the great advantages of a mesoscale model is that it can be used to test arbitrary new outputs, allowing the model to be falsified and updated in an iterative fashion.

\label{sec_input_and_output_specification}
\section{Model Overview}

In this section, we describe what researchers are required to provide as input in order run an existing model and what they can expect as output. We close the section with a high-level overview of how all the pieces fit together in preparation for more detailed discussion in subsequent sections.


\subsection{Model Input}


A mesoscale model requires as {\it{input}} a source of structural data which in our case corresponds to an annotated connectome. By connectome I mean the equivalent of a 3D reconstruction of all the neuropil within the target tissue, which could be the complete brain of an organism or a subcircuit such as the optic lobe of the fly. Practically speaking, we are interested in the fly brain, 100,000 neurons and approximately 10 million synapses, and the similarly-sized brain of the larval zebrafish.

Regarding the connectome, we are interested in the graph consisting of neurons as vertices and edges that represent a suitable summary of the connections between pairs of neurons. In our case we are interested in what is formally called a {\it{multi-graph}} in which there is a directed edge from $A$ to $B$ for each synapse such that $A$ is the pre-synaptic neuron and $B$ is the post-synaptic neuron. It's important to note that the adjacency matrix by itself is not sufficient for our purposes, the connectome graph is embedded in a 3D volume whose geometry approximates the spatial properties of the original tissue. In particular, we require a spatial database that includes the coordinates of every neuron and every synapse within the tissue. In addition we require cell types, morphology and synapse metadata for calculating estimates of connection strength such as vesicle counts and synaptic-cleft measurements.

To infer functional models\footnote{%
  The basic problem of inferring the function of a neural circuit is sometimes called {\it{functional connectomics}} {\cite{AlivisatosetalNEURON-12,ReidNEURON-12,SeungNATURE-11}} in recognition of David Hubel and Torsten Wiesel's use of the term {\it{functional architectures}} in describing the relationship between anatomy and physiology in cortical circuits~\cite{HubelandWieselJoP-68,HubelandWieselJoP-62}.},
we require a source of functional data. In this case, we assume state-of-the-art methods for recording from all of the neurons in the target tissue, such as the use of genetically-modified organisms that express fluorescent proteins indicative of functionally-relevant activity and two-photon-excitation microscopy to record estimates of local calcium flux as a proxy for electrical activity. The result is a time series of point clouds, where each point cloud is embedded in a 3D volume the same size as the connectome graph embedding. Within a point-cloud embedding volume, each {\it{point source}} represented in the cloud corresponds to the fluorescent emissions emanating from a single neuron cell body, generally confined to the cell nucleus or the area adjacent to the axon hillock. 

Since we are attempting to bridge the gap between the (microscale) cellular and the (macroscale) behavioral scales, we need a source of behavioral data. For now, suffice it to say that we expect to acquire high-speed video of the target as part of the experimental protocol. Both functional and behavioral data contain timestamps to facilitate alignment. Since the raw footage is bulky and inadequate as ground truth without additional annotation, researchers working on organisms like \Celegans{}, drosophila and zebrafish perform automated analyses of the video to identify stereotypical micro-behaviors referred to as {\it{behavioral modes}}. For example, a simple reaching behavior may have preparation, initiation, extension and termination modes. Using the resulting mathematical characterization, it is then possible to decompose complex behaviors into their component modes, so as to more precisely align neural state changes with behavior.

Finally, we assume we are given an {\it{inductive bias}} that constrains the class of models and facilitates generalization. There are only two types of inductive bias that can claim theoretical justification for their ability to generalize from observed instances of a given class to instances not appearing in the training data. They are {\it{model minimization}} as in the application of the minimum-description-length principle~\cite{Rissanen78} and {\it{margin maximization}} exemplified by support vector machines~\cite{CortesandVapnikML-95}. In our case we use the method of minimizing models to focus attention on a restricted class of configurable neural networks that are used to instantiate what we refer to as {\it{functional modules}}. In the literature, the term functional module typically refers to relatively small circuits purported to exhibit modular function. We extend the term to apply individually to each subcircuit in a disjoint set of subcircuits that together comprise a single connected circuit representing the entire target tissue. For our purposes, a functional module consists of a domain, a configurable network and an {\it{interface}} that serves much the same purpose as an application program interface or API in software engineering.


\subsection{Model Output}


A mesoscale model defines as {\it{output}} a partition of the recorded point sources into possibly overlapping functional domains, where we use the word domain as it is generally used in mathematics to represent the domain of a function but then occasionally abuse the terminology to mean the set of all inputs and outputs, i.e., the domain and range, in discussing properties of graphs and graph algorithms. 

Each point source represents information relating to the {\it{neural state}} of its corresponding neuron. The mesoscale model maintains a {\it{neural state vector}} that tracks the state of all the neurons in the microcircuit. It is the job of the functional modules, or more precisely their configurable networks, to infer the neural state vector at $t+1$ from the neural state vector and the corresponding vector of point sources at $t$.

The mesoscale model is recurrent since it maintains this neural state vectors. In contrast, it is a design goal that the configurable neural networks that instantiate functional modules are pure functions\footnote{%
  A pure function always evaluates to the same output given the same input. Strictly speaking a pure function can't depend on any hidden state that changes during evaluation or between separate executions. Since most neural computations are stateful and recurrent, we side-step the strict definition by assuming the recurrent state is always provided as one of the inputs.}, 
as this requires that the only recourse for saving state is to save it in the neural state vector. One alternative we've considered and will touch upon later involves adding hidden layers that attempt to infer the status of individual synapses.

In addition to defining the domains for each functional module we must also map each functional module to a configurable neural network. Here we separate the identification of functional domains from the assignment of configurable networks, but it should be apparent that all three components of a functional module including its domain, network and interface are inextricably dependent on one another and therefore must be inferred jointly.

Finally, a mesoscale model defines a sparse functional basis that serves to explain all the computations performed in each domain. Since this concept may be unfamiliar to some readers, I will unpack the concept with some care in subsequent sections as it is one of the key features of the proposed class of mesoscale models, providing our ability to identify the distribution of functional motifs occurring throughout the target tissue.

For the reader familiar with sparse coding, a sparse functional basis is simply a basis for a function space rather than a vector space as is generally assumed in sparse coding. Analogous to sparse coding, a sparse functional basis corresponds to a set of basis functions that define a family of functions in which each function in the family can be defined as a linear combination of the basis functions such that only a small number of the weights that define the linear combination are nonzero.


\subsection{Basic Model}

\begin{figure}
  \begin{center}
    \includegraphics[scale=0.185]{\figures/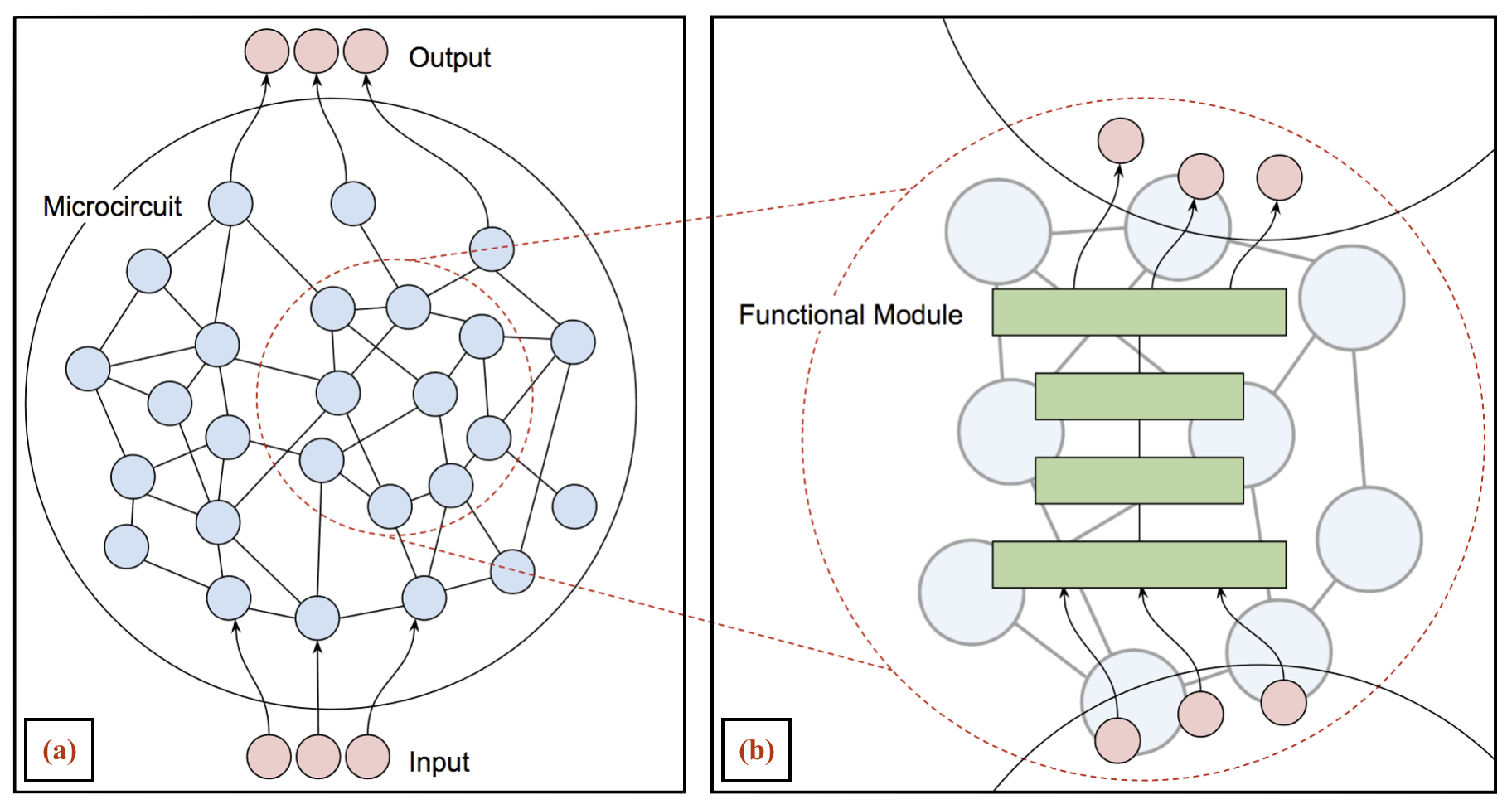}
  \end{center}
  \caption{The graph shown in the large circle\emdash{}outlined in solid black\emdash{}on the left (a) represents the full microcircuit connectome graph and the enclosed circle—outlined in dashed red—represents the domain of one functional module. The inset on the right (b) depicts the functional network associated with the inscribed module. The mesoscale model combines the functional networks associated with the model functional modules into one large recurrent network. In the simplest arrangement, this composite network takes sensory patterns as input and produces activity patterns as output. The mesoscale model loss function includes two terms relating to prediction accuracy: One term measures how well the model as a whole reproduces the recorded activity given the associated sensory input. The other term measures the ability of the individual functional networks to reproduce the activity observed in their respective functional interfaces. The second term is offset by a third term in the loss function that penalizes the complexity of functional networks calculated as a function of the number, size and type of their component layers. The combined second and third terms constitute a proxy for explanatory value.}
  \label{fig_training_modules_microcircuits_architecture}
\end{figure}

To get some perspective before we discuss the architectural details, consider the schematic rendering of the basic model shown in Figure~\ref{fig_training_modules_microcircuits_architecture}.(a). The graph enclosed in the circle labeled "microcircuit" represents the neural circuitry of the target organism which, owing to the limitations of current state-of-the-art recording technologies, might, in the near term, correspond to a fly, zebrafish or some interesting subcircuit such as the visual or olfactory system of a fly or the striate cortex in mammals. 


The "input" can be thought of as the stimuli that the organism is exposed to in the process of a researcher carrying out an experimental protocol. The sensory environment for a small organism is likely to include chemosensory, thermosensory and visual stimuli as well as precisely targeted optogenetic stimulation of individual neurons and neuronal groups.

The "output" corresponds to the observed behavior. Typically, behavioral data is acquired using high-speed cameras generating thousands of frames per second. The resulting video is then analyzed to generate a more compact, mathematically precise account of behavior, often by using Fourier analysis and spectral-graph algorithms to reveal microscale behavioral modes that allow a more detailed account in keeping with the temporal resolution of the functional recordings.

In addition to inputs and outputs that relate the macroscale behavior of the whole organism to its external environment, we are also interested in the microscale inputs and outputs of smaller neural circuits available from the functional data as measurements of calcium flux or local field potential. Recordings of neural activity at cellular resolution enable us to evaluate proposals for decomposing large circuits comprised of many neurons into modular networks comprised of relatively few components, each one accounting for a subset of the neurons in the larger circuit.

The graphic shown in Figure~\ref{fig_training_modules_microcircuits_architecture}.(b) illustrates the idea of a functional module that accounts for a subcircuit of the full microcircuit. Each such module has a {\it{domain}} that determines its inputs and outputs, an {\it{interface}} that accounts for its functional relationships with other modules, and an artificial neural network that defines it's implementation. Each network corresponds to an instance of a class of {\it{configurable neural networks}}.

We refer to these configurable neural networks as {\it{shallow deep neural networks}}. They are {\it{deep}} in that they utilize the same basic component layers found in the current generation of deep neural networks, and {\it{shallow}} in that they are comprised of a small number of components layers well known to most neuroscientists because they were either discovered by neuroscientists studying the brain or inspired by the work of neuroscientists. Examples include divisive normalization, max pooling, and the sigmoidal, logistic and rectified-linear units often employed in hidden layers.

The motivation for using shallow deep-neural-network architectures is that functional-module networks are intended to serve as computational explanations of local neural function in our model. Given that we understand each of the component layers, it is relatively easy to explain what's going on in a shallow network in much the same way that a short computer programs written in a conventional programming language is easy to understand if one steps through the program line by line. This is true even if the program manifests complex behavior, as in the case of a short program implementing the one-dimensional logistic map that manifests chaotic behavior.

While we have some way to go in mathematically understanding the unreasonable effectiveness of deep artificial neural networks, software engineers and machine-learning researchers who routinely work with such networks can design, test, debug and explain deep-neural-network architectures constructed from standard components. At least in the case of shallow networks consisting a relative small number of layers, such adepts can provide a reasonably intuitive characterization of their behavior.


\label{sec_basic_technology_mathematics}
\section{Background}


Since this paper is intended for researchers from diverse backgrounds and the model and method of learning it are sufficiently complicated, we'll take a few minutes here to review the relevant terminology, technology and mathematical concepts. Many of the concepts are found in introductory courses on linear systems theory and functional analysis. Some readers may be familiar with the ideas from developing programs in languages that support matrix-vector operations for applied linear algebra or from learning about the widely-used and unreasonably-successful class of artificial neural networks known as convolutional neural networks. To understand how we intend to learn and then apply a sparse functional basis in a mesoscale model, it will help if you are somewhat familiar with the terminology employed in working with linear filters, convolving a mask or kernel matrix with an image, training convolutional neural networks, learning a sparse basis and using it to reconstruct a noisy signal, and understanding how every continuous function in a function space can be represented as a linear combination of basis functions.


\subsection{Filters and Convolutional Networks}



\begin{figure}
  \begin{center}
    \includegraphics[scale=0.195]{\figures/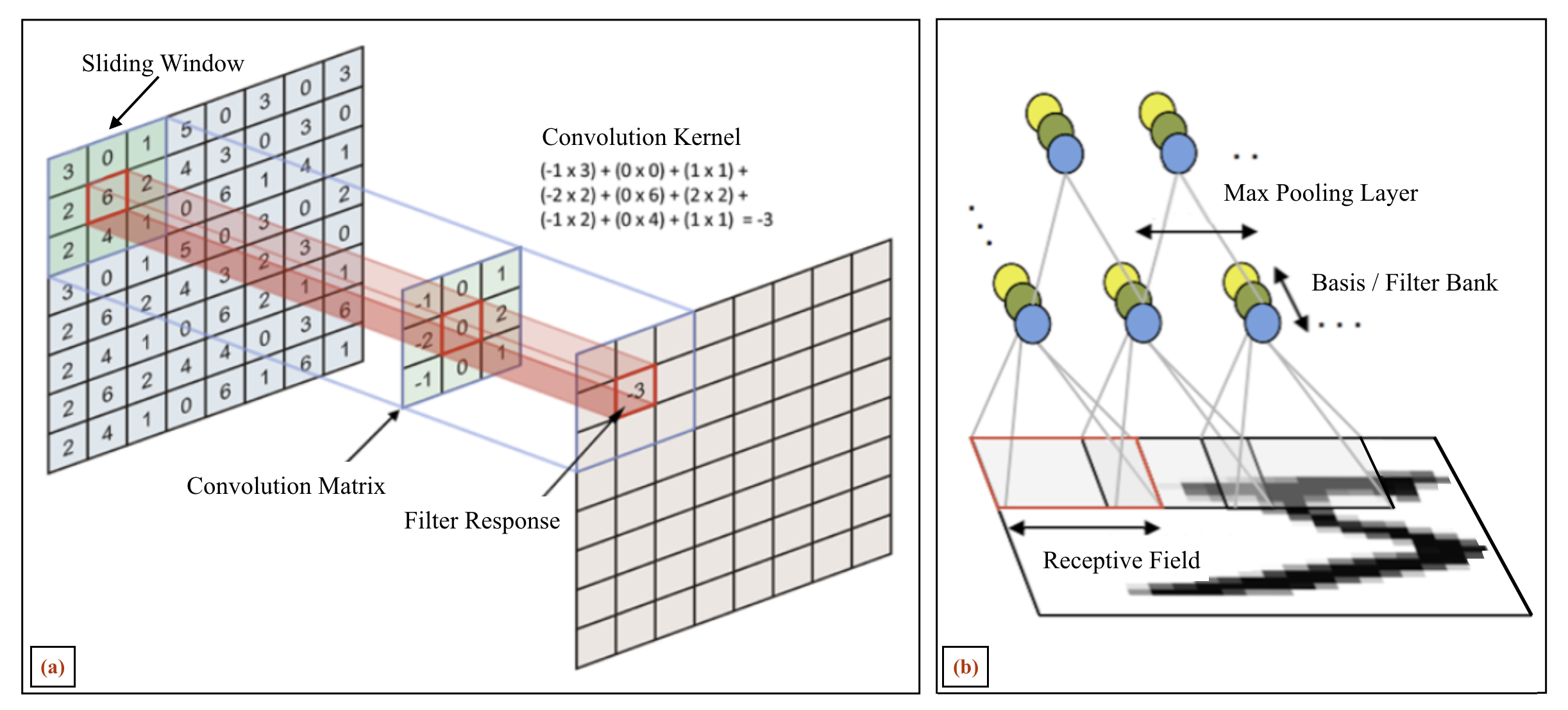}
  \end{center}
  \caption{The graphic on the left (a) illustrates the basic elements involved in convolving a $3 \times{} 3$ filter with a $8 \times{} 8$ matrix to compute a response matrix. The graphic on the right (b) depicts a typical application of convolution in the context of convolutional neural networks in which each filter in a bank of $B$ filters is convolved with an $H \times{} W$ image resulting in a $H \times{} W \times{} B$ stack of responses which is then convolved with a max pooling kernel.}
  \label{fig_000_convolutional_architectures_sparse_coding_and_max_pooling}
\end{figure}


A simple gray-scale image is generally implemented as a two-dimensional matrix. However, when an image is encoded as the input layer in an artificial neural network, it represents a vector in a vector space of dimension equal to the number of pixels in the image. When dealing with high-dimensional inputs such as images, it is impractical to connect all the units in the input layer to all of the units in, for example, a subsequent hidden layer.

Instead, we only connect each unit in the hidden layer to a local region of the input layer. The spatial extent of this connectivity is a hyperparameter called the {\it{receptive field}} or {\it{filter size}} of the units in the hidden layer. Substitute the word "neuron" for "unit" and the same terminology is applied to neural circuits, especially circuits organized as topographic maps such as are found in the retinotopic maps appearing in the ventral visual stream of mammalian neocortex.

The size of the hidden layer is determined by the receptive field and the spacing or {\it{stride}} between the local regions assigned to hidden units. In image processing, a {\it{kernel}} or {\it{convolution matrix}} is a small matrix that serves as a {\it{filter}} on the input image used for edge detection, image blurring and other transformations. These transformations are accomplished by means of a convolution between a kernel and an image. A collection of filters is called a {\it{filter bank}}.

The term {\it{filter kernel}} or {\it{kernel function}} is generally employed when discussing the obvious generalization of convolution. In many image processing tasks, the kernel function is the dot product of the convolution matrix and a filter-sized region of the target data, corresponding to a matrix, volume or other form of structured data. In a convolutional neural network the entries in the convolution matrix are free parameters that are learned via back propagation.

When computing the convolution of an image with a filter, we run a {\it{sliding window}} over the image, applying the kernel function to the kernel at each receptive-field-sized local region of the input layer as defined by the filter size and stride parameters, resulting a filtered image the same size as the hidden layer. The coordinates of the local regions constitute a grid spanning the input layer, a 2D grid in the case of simple gray-scale images or 3D grid in the case of mesoscale volumetric data.


\subsection{Learning Sparse Representations}


Learning a {\it{complete}} basis for representing a class of objects $X$ such as images that can be encoded as vectors involves searching for a set of basis vectors $B = \{ \phi_{i} | 1 \leq i \leq K \}$ such that we can represent {\it{any}} element $x \in X$ as a linear combination of these basis vectors, in which case $B$ is said to {\it{span}} $X$. The {\it{reconstruction error}} is the absolute value of the difference between $x$ and its representation as a linear combination of basis vectors is shown in Equation~\ref{eqn_state_reconstruction_error}. 
\begin{equation}\label{eqn_state_reconstruction_error}
   \left\|x - \sum_{i=1}^{K} \alpha_{i} \phi_{i} \right\|
\end{equation}

Unlike Principal Components Analysis (PCA) that enables us to learn a complete set of vectors, sparse coding learns an {\it{over-complete}} set of basis vectors that are better able to account for interesting structure in the input. In the case of an over-complete basis, the coefficients $\alpha_{i}$ are no longer uniquely determined by the input and hence we introduce the requirement of sparseness to resolve the degeneracy introduced by over-completeness. The loss function for learning a sparse basis is a weighted combination of the sum of the reconstruction error over the set of $m$ training examples and a sparsity-inducing penalty function as shown in Equation~\ref{eqn_sparsity_inducing_penalty}.
\begin{equation}\label{eqn_sparsity_inducing_penalty}
   \mbox{\rm{minimize}}_{(\alpha_{i}^{(j)},\phi_{i})} 
   \sum_{j=1}^{M} \left\| x^{(j)} - 
     \sum_{i=1}^{K} \alpha_i^{(j)} \phi_i \right\|^2 + \lambda \sum_{i=1}^{K} S(a_i^{(j)})  
\end{equation}

The penalty function is generally based on various norms, similar to those employed in compressive sensing. The $L_0$-norm is the number of non-zero elements in a vector and hence is integer-valued and not differentiable, so optimization is often carried out by some variant of the analysis-synthesis method for solving convex-optimization problems~\cite{AroraetalPMLR-15}. The $L_1$-norm, also known as {\it{least absolute deviations}}, can be used with gradient methods and consists of minimizing the sum of the absolute differences between the target and the reconstruction of the target as a linear combination of the basis vectors.

The set of basis vectors learned by sparse coding can be thought of as a code book or dictionary and is employed as such in applications of the theory of compressive sensing and sparse representation for object recognition, audio blind source separation and super-resolution image enhancement~\cite{PatelandChellappa2013,PatelandChellappaACPR-11,PlumbleyetalIEEE-10}. The basic idea for super-resolution image enhancement is that, having trained a sparse basis on high-resolution images of the same sort you want to enhance, you divide the low-resolution target image into patches, reconstruct each patch as a sparse linear combination of the high-resolution basis vectors and then perform an image-wide optimization to create a composite high-resolution image~\cite{PatelandChellappa2013,YangetalTIPS-10}. 

\subsection{Basic Functional Analysis Concepts}

We don't require any complicated concepts from functional analysis besides the notions of {\it{function space}} and {\it{function basis}} analogous to vector space and vector basis. We define the function space $\hmFs$ as the space of all linear combinations of finite subsets of a set $\C$ of continuous vector-valued functions, $\{ f : \R^{M} \rightarrow \R^{M} \}$, that are realized as configurable neural networks. The functions in $\hmFs$ will serve to map the neural state vector at time $t$ to the neural state vector at time $t+1$.

We propose to search this space in order to learn a sparse basis that spans the space of emergent computational motifs occurring at different scales within a given neural microcircuit. We hypothesize that such motifs occur and will be recognizable in multiple locations within individual phenotypes, across individuals of the same genotype and perhaps across species, genus or more distantly related taxonomic ranks. 

In order to instantiate each configurable network, we need to specify the number of layers, and for each layer determine its activation function, number of units, and the connection pattern of its weight matrix. In designing a network by hand, these choices are discrete, e.g., involve a binary choice or selection from a fixed menu of options.

Since in the approach described in this paper we have to make such choices by back-propagating the gradient of a loss function, all parts of the model architecture involved in configuration have to be differentiable, and so, instead of a conventional discontinuous step function, we rely on the logistic function which is a smooth approximation of the unit (Heaviside) step function\footnote{%
  Alternatively, we could apply reinforcement learning to inferring a mesoscale model by using the loss function as a reward signal and a set of possible actions that correspond to setting network configuration parameters to discrete values, e.g., four hidden layers of 100 units each with a fully-connected (bipartite) weight matrix. Zoph and Le~\cite{ZophandLeICLR-17} use such a strategy to learn network architectures in which actions are strings constructed from a grammar and lexicon for specifying network parameters.}

In addition to setting network configuration parameters, there are two other sets of choices that ideally require a binary or one-of-$N$ choice, but we have to make do with a less decisive, differentiable alternative. They involve (a) input assignments mapping each component of the neural state vector at time $t$ to exactly one basis function, and (b) output assignments selecting the basis function responsible for predicting the $i$th component of the neural state vector at time $t+1$. Learning such assignments is algorithmically equivalent to solving a graph partition problem~\cite{GareyandJohnson79}.

Strictly speaking the functions in $\C$ are defined on a subspace of $\R^{M}$ representing the (vector) space of all possible neural states. In practice, their respective domains and ranges are restricted to subspaces that correspond to localized populations of neurons. Part of defining a functional basis that represents the dynamics of a neural microcircuit involves learning not just the configuration of basis functions realized as neural networks, but also their restricted domains and ranges. 

Suppose we have a functional basis $\B = \{ \beta_1, \beta_2, \ldots, \beta_{|\B|} \}$ where each basis function $\beta_i$ corresponds to the configurable network of a functional module and is responsible for accounting for the dynamics of some subcircuit of the neural microcircuit. Learned input assignments determine the domains of each basis function, $\Domains{} = \{ D_1, D_2, \ldots, D_{|\B|} \}$ and corresponding output assignments determine their ranges $\Ranges{} = \{ R_1, R_2, \ldots, R_{|\B|} \}$, so that $\beta_{i}: D_{i} \rightarrow R_{i}$ for $i = 1,\ldots,|\B|$. 

\begin{equation}\label{eqn_sparse_linear_combination}
  F(\mathbf{x}^{t}) = \mathbf{x}^{t+1} = \sum_{i = 1}^{|\mathcal{B}|} \mathbf{w}\big|_{\mathcal{R}_{i}} \beta_{i}(\mathbf{x}^{t}\big|_{\mathcal{D}_i})
\end{equation}

Let $\mathbf{v}$ denote a vector and $\mathbf{v}\big|_{S}$ denote a restriction or {\it{mask}} that allows the basis function to operate on a subspace $S$, thereby updating precisely defined portions of the neural state vector. Equation~\ref{eqn_sparse_linear_combination} illustrates how we might update a neural state vector as a linear combination of range- and domain-restricted basis functions. 

Unfortunately, this approach won't work for the complex neural circuits we are interested in. Equation~\ref{eqn_sparse_linear_combination} completely misses the most important characteristics of the problem we are faced with in this paper, namely, the spatial structure and local geometry of neural circuits.  A good deal of the sequel is devoted to representing and exploiting the physical structure of real neural networks.

As we are advocating a rather complicated architecture, it is worth considering if a simpler approach might suffice. The universal approximation theorem for artificial neural networks tells us that, theoretically, we need only one basis function whose receptive field is the entire neural state vector, and that function could be defined by a multi-layer perceptron with only one hidden layer\footnote{%
  Let $\F$ be the space of all continuous functions defined on $\R^{M}$. The universal approximation theorem for artificial neural networks tells us that $\forall{} f \in \F$, and $\forall{} \epsilon > 0$, there exist real constants $\upsilon_i, b_i \in \R$ and real vectors $w_i \in \R^{M}$, where $i = 1,\ldots,N$, such that we may define:
  \begin{equation}
    F(x)=\sum_{i=1}^{N} \upsilon_i \sigma(w_{i}^{T}x + b_i)
  \end{equation}
  as an approximation of $f$ in the sense that:
  \begin{equation}
    \left\|F(x) - f(x)\right\| < \epsilon
  \end{equation}
  for all $x \in \R^M$~\cite{CybenkoMCSS-89,HorniketalNEURAL-NETWORKS-89}. Here we stipulate the logistic activation function $\sigma{}$, but the theorem holds under reasonable conditions for any non-constant, bounded and monotonically-increasing continuous function~\cite{CsajiMSc-01,HornikNEURAL-NETWORKS-91}.}.
This might be a reasonable strategy if you could exploit the connectome graph to avoid a combinatorial explosion in the number of weights in the hidden layer. However, by the time you did figure out how to simplify the hidden layer enough to make the implementation tractable, you would have done as much work as we are proposing to do here by learning a sparse basis.

However, the main disadvantage of learning a single monolithic network is that it does not forward our stated goal of understanding biological computation at the mesoscale. By demonstrating we can learn a sparse functional basis for a complex microcircuit, we stand a chance of identifying computational strategies that appear repeatedly within an individual, that arise in a variety of evolutionarily conserved regions and are found across different species. 






\label{sec_model_architecture}
\section{Model Details}


In this section, we describe the conceptual and technical interplay between the class of mesoscale models introduced in this article and the network architecture used to learn such models. In particular, we describe some of the most important technological desiderata leading to the design of the underlying network architecture. This exercise offers an extended example illustrating the richness of artificial neural networks and their representational value in terms of explaining biological computation.


\subsection{Constructing  Connectomic Graphs}


\begin{figure}
  \begin{center}
    \includegraphics[width=\textwidth]{\figures/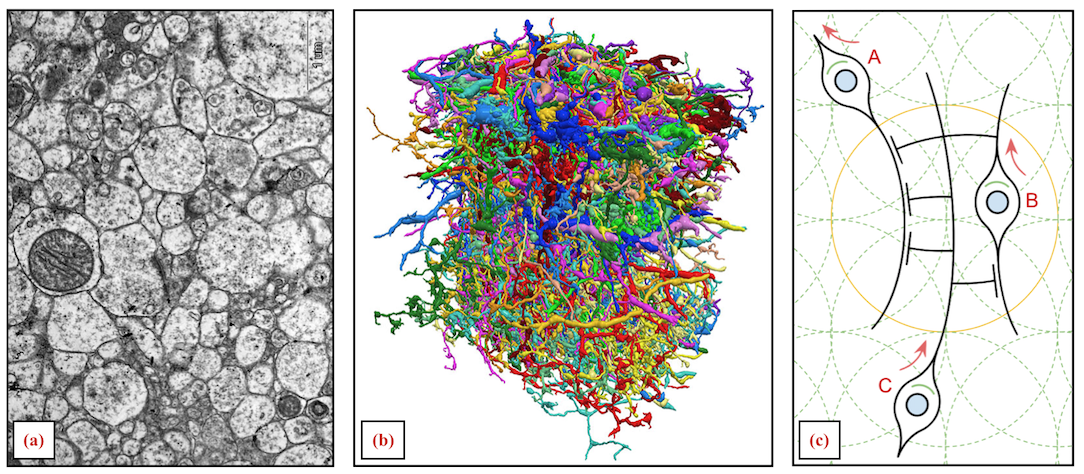}
  \end{center}
  \caption{This slide and the next describe the sources and procedures for combining structural and functional data to create the time series necessary to train mesoscale models. Panel (a) shows detail from an electron micrograph (EM) of a cross section of neural tissue obtained using a diamond knife to repeatedly shave off the top few nanometers from a block of preserved tissue. Thousands of such sections are required to achieve sufficient resolution to compute a dense reconstruction (b) of all the neuropil in a sample 100 microns on a side.}
  \label{fig_001_structural_connectomics_and_microcircuit_connectome_graphs}
\end{figure}


Figure~\ref{fig_001_structural_connectomics_and_microcircuit_connectome_graphs}.(a) shows a sample electron micrograph of the neural tissue used to construct the connectome graph. The dense packing of neuropil is clearly evident in the micrograph. The full structural dataset consists of a stack of such sections that together comprise a 3D representation of the target tissue. The resolution varies depending on the imaging technology. Transmission (TEM) and scanning (SEM) electron microscopy using diamond knife serial sectioning produce anisotopic imagery with $\approx{}10$nm in $x$ and $y$ and $\approx{}$20nm in $z$. Focused ion beam (FIB) scanning EM ablates the surface of a prepared tissue block thereby achieving approximately isotopic imagery with $\approx{}$5-10nm in each of $x$, $y$ and $z$. Current state-of-the-art EM technology is a lot more varied and complicated that suggested here\emdash{}see Marblestone~\etal~\cite{MarblestoneetalBIORXIV-14} for a more comprehensive review.

Figure~\ref{fig_001_structural_connectomics_and_microcircuit_connectome_graphs}.(b) shows a dense reconstruction of all the neuropil in a tissue sample including the complete elaboration of the axonal and dendritic arbors. A reasonably complete, well annotated and curated connectomic dataset includes skeletons, adjacency matrices, and features relating to synaptic weight and connection strength, morphology, cell types, that can be inferred from the EM imagery. For example, it is often possible to estimate the distance across the synaptic cleft separating the pre- and post-synaptic neuron and the number of neurotransmitter vesicles to estimate the strength and valence\emdash{}excitatory or inhibitory\emdash{}of each synapse.


Figure~\ref{fig_001_structural_connectomics_and_microcircuit_connectome_graphs}.(c) depicts a simple cartoon of a neural circuit with three neurons and four shared synapses. The four synapses and one of the three neuron cell bodies are enclosed in a circle representing a cross section of a spherical subvolume of the microcircuit connectome embedding space. The four synapses and their corresponding pre- and post-synaptic neurons constitute a subgraph of the full connectome. It is this subgraph that will serve as a starting point for defining the domain of a given functional module.

There are several ways we might define a connectome graph $G = (V,E)$. If the set of vertices and the set of neurons are in a one-to-one correspondence, there are two alternatives we can consider here. In the first alternative, there is a exactly one edge from $i$ to $j$ in $V$ if and only if there exists 1 or more synapses such that $i$ corresponds to the presynaptic neuron and $j$ to the postsynaptic neuron. Alternatively, $G$ is a multi-graph such that there are the same number of edges from $i$ to $j$ as there are distinct synapses such that $i$ corresponds to the presynaptic neuron and $j$ to the postsynaptic neuron.

Since we currently can't record from all of the synapses in any reasonably large tissue, synapses remain implicit, their weights corresponding to hidden variables that we attempt to learn from a combination of neural circuitry available in the connectome and functional recordings of their pre- and post-synaptic neurons. This structural and functional information could allow us to infer synaptic strength given that we record from all of the neurons simultaneously; however, more likely than not, the problem is significantly under constrained and the best we can hope for is a rough estimate. 


\subsection{Integrating Structure and Function}


\begin{figure}
  \begin{center}
    \includegraphics[width=\textwidth]{\figures/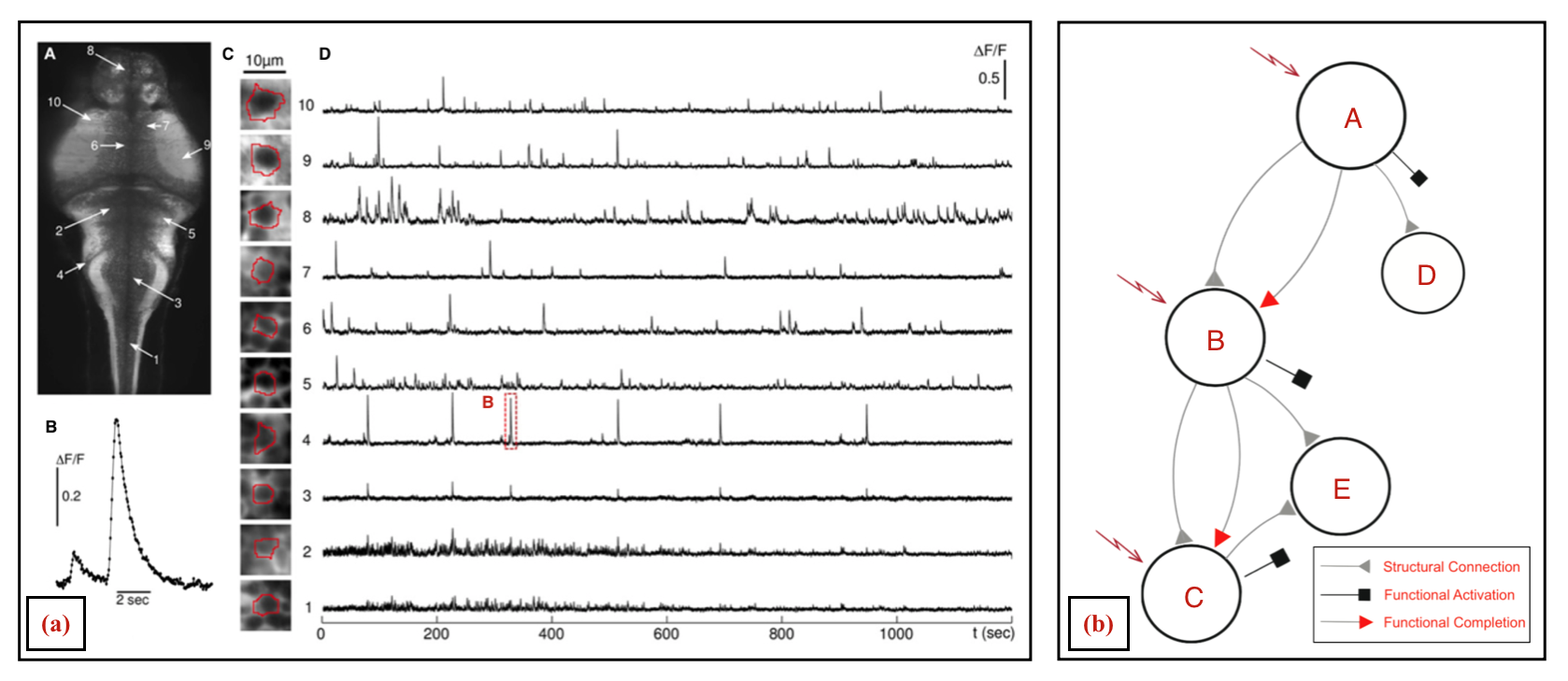}
  \end{center}
  \caption{Panel (a) describes a functional dataset from a larval zebrafish showing (\redhighlight{A}) a time-averaged image of a slice of the zebrafish brain obtained from two-photon microscopy showing the location of 10 neurons, (\redhighlight{B}) a detail plot of the change in fluorescence $\hmDelta{}F/F$ of a single neuron as a function of time, and (\redhighlight{C}) thumbnail images of each neuron's region of interest (integration) and the corresponding time-series called a {\it{raster}} spanning the single-trial interval\emdash{}adapted from Figure~3 in Panier~\etal~\cite{PanieretalFiNC-13}. Panel (b) shows a simple neural circuit referred to in the text to explain the transmission response paradigm and its application to combining the structural and functional data\emdash{}adapted from Figure~4 in Dlotko~\etal{}~\cite{DlotkoetalCoRR-16}.}
  \label{fig_002_functional_connectomics_and_transmission_response_graphs}
\end{figure}




  


Figure~\ref{fig_002_functional_connectomics_and_transmission_response_graphs}.(a) describes a functional dataset recorded from a larval zebrafish. The most important characteristic of this dataset for our immediate purpose is that it provides the 3D coordinates of each recorded neuron and an estimate of the calcium flux in (typically) the cell nucleus at each point in time throughout the single-trial interval, that can be used as a rough proxy for neural activity / local field potentials.. 

The graphic in Figure~\ref{fig_002_functional_connectomics_and_transmission_response_graphs}.(b) illustrates the {\it{transmission-response paradigm}} that we use to combine structural and functional data in order to capture the model dynamics in a series of {\it{transmission-response graphs}} each of which is a subgraph of the connectome graph with all of the vertices in the connectome graph and a subset of the edges. The graphic shows a toy connectome graph. We assume that all and only those synapses present in the microcircuit are registered as edges in the connectome graph. We construct the sequence of transmission-response graphs as follows:

Divide the interval over which the functional data is collected into subintervals of some fixed duration. For each subinterval $p$, construct a transmission response graph $G'= (V, E')$ such that, for each edge $e$ in the original connectome graph $G = (V, E)$ , $e$ is an element of $E'$, if and only if the functional data collected over the interval $p$ supports the hypothesis that an action potential is propagated across the corresponding synapse from the pre- to the post-synaptic neuron during $p$. So in the case of Figure~\ref{fig_002_functional_connectomics_and_transmission_response_graphs}.(b), for the relevant $5$ms interval $i$, each of the three neurons \redhighlight{A}, \redhighlight{B}, and \redhighlight{C} spike in the interval $i$, \redhighlight{B} spikes within $7.5$ms of \redhighlight{A}, \redhighlight{C} spikes within $7.5$ of \redhighlight{B} and only the edges \redhighlight{A} $\rightarrow$ \redhighlight{B} and \redhighlight{B} $\rightarrow$ \redhighlight{C} are marked as functionally completed and thus included in the transmission response graph associated with the $i$th interval.


\begin{figure}
  \begin{center}
    \includegraphics[width=\textwidth]{\figures/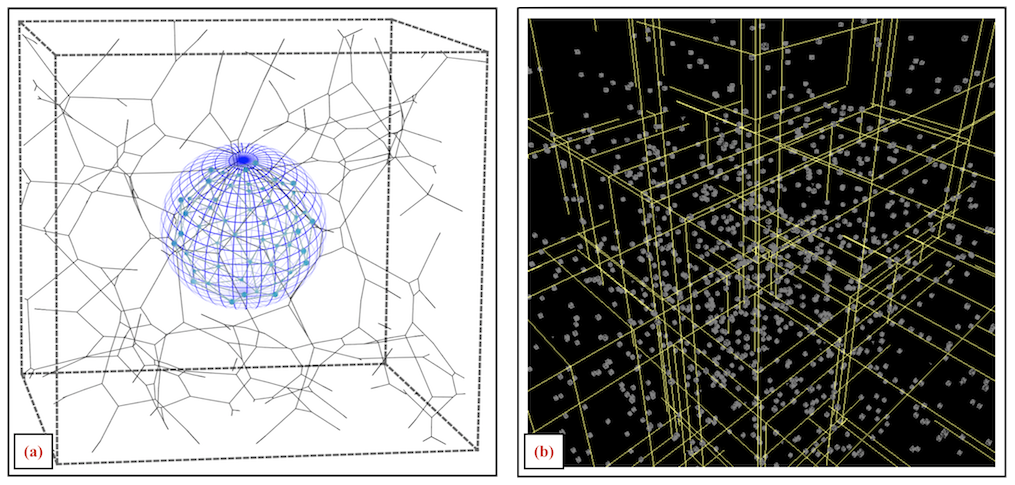}
  \end{center}
  \caption{Panel (a) depicts a microcircuit connectome graph embedded in a 3D volume and a spherical subvolume corresponding to a convolution receptive field. To efficiently train our mesoscale models, we need to be able to extract the subgraph whose synapses fall within such 3D volumes and do so quickly and in parallel. Panel (b) depicts a KD tree optimized for fast nearest neighbor search and subvolume extraction.}
  \label{fig_003_spill_trees_convolution_filters_and_spherical_receptive_fields}
\end{figure}


Figure~\ref{fig_003_spill_trees_convolution_filters_and_spherical_receptive_fields}.(a) depicts a microcircuit connectome graph embedded in a 3D volume and a spherical subvolume corresponding to a convolution receptive field. Applying the terminology of convolutional neural networks, a functional basis corresponds to a set of functional modules, the kernel for a given functional basis filter corresponds to the configurable network associated with the module, the free parameters of the kernel correspond to the configuration parameters that determine a fully instantiated configurable neural network. The kernel function applies this network to functional domains extracted from the local regions.

The input layer of a mesoscale model is a 3D embedding of a transmission response graph, not a simple matrix or higher-dimension tensor. The grid traversed by the sliding window operator is defined by a set of 3D points spanning the volume corresponding to the neural microcircuit. The local regions corresponding to receptive fields are spherical subvolumes of a fixed diameter separated with a stride substantially less than the diameter. The sliding window operator is implemented by a highly-parallel distributed spatial index based on KD trees as illustrated in Figure~\ref{fig_003_spill_trees_convolution_filters_and_spherical_receptive_fields}.(b). To expedite training, we employ infrastructure consisting of multiple distributed servers optimized to handle many requests in parallel. 


\subsection{Convolutional Network  Architecture}


\begin{figure}
  \begin{center}
    \includegraphics[width=\textwidth]{\figures/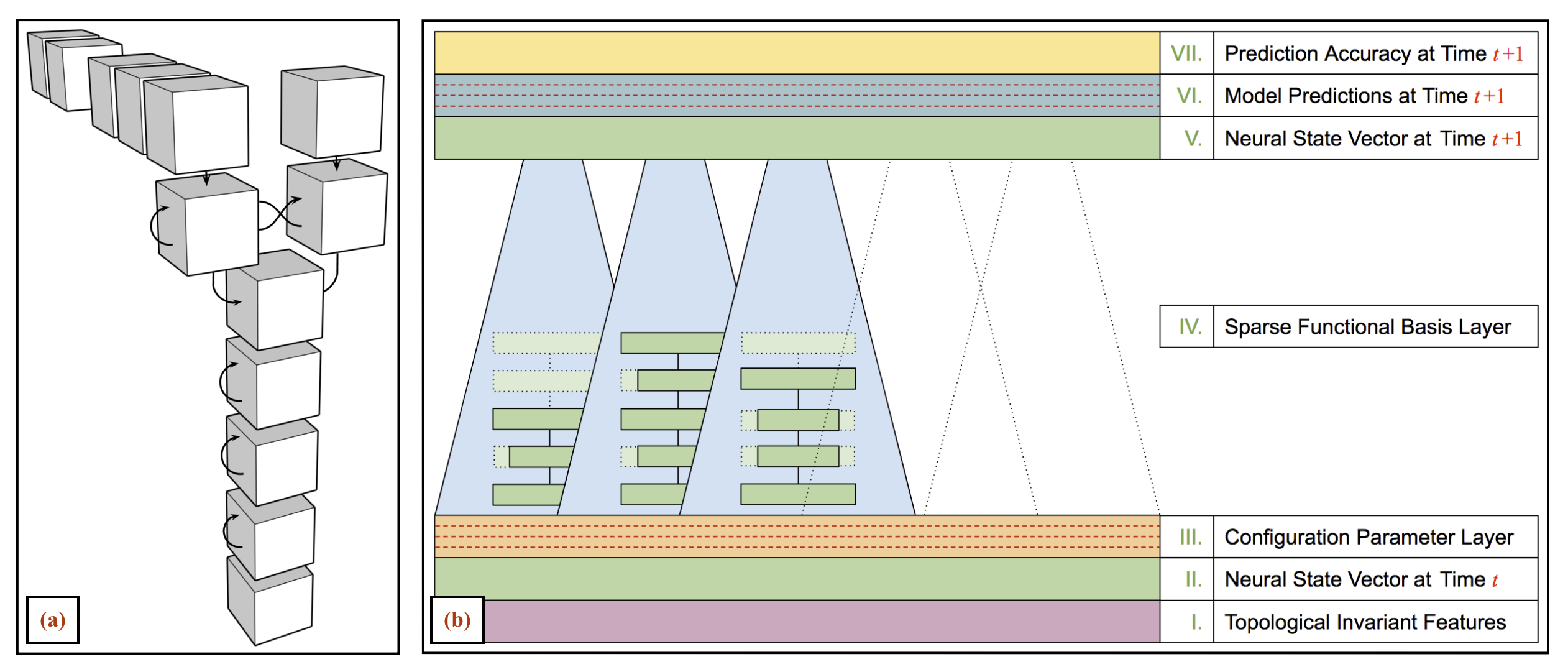}
  \end{center}
  \caption{The graphic on the left (a) depicts the architecture of the convolutional neural network implementing the mesoscale model showing how the static structural (connectome) data (top right) is fused with each 3D slice of the dynamic (two-photon) functional data to generate the sequence of transmission response graphs illustrated in Figure~\ref{fig_002_functional_connectomics_and_transmission_response_graphs}. The diagram on the right (b) represents seven layers of the neural network architecture for learning a mesoscale model. These seven layers illustrate only a part of the complete architecture and are primarily concerned with learning the sparse basis in the form of a set of functional modules, including their associated configurable network and interface.}
  \label{fig_003_mesoscale_model_convolutional_neural_network_architecture}
\end{figure}


Figure~\ref{fig_003_mesoscale_model_convolutional_neural_network_architecture}.(a) represents an inverted architecture for a convolutional neural network with input on the top and output on the bottom in contrast to conventional renderings. The cube on the top right is the the 3D embedding of static connectome graph, providing a topographical map of the neural tissue. The sequence of cubes on the top left represents the time series of point clouds embedding the two-photon-excitation fluorescent point sources with the same geometry as the static connectome embedding.

The penultimate layer\emdash{}second from the top in Figure~\ref{fig_003_mesoscale_model_convolutional_neural_network_architecture}.(a)\emdash{}fuses the static structural data and the dynamic functional data to generate the series of transmission response graph embeddings described in Figure~\ref{fig_002_functional_connectomics_and_transmission_response_graphs}. The remaining layers ingest each transmission response graph in sequence updating the neural state vectors, and integrating new data to estimate the error in predicting the next neural state vector.

The layers following the functional-structural fusion layer are responsible for adjusting the {\it{global}} functional module network configuration parameters as well as the parameters that define the {\it{local}} functional module interface layers and assign point sources to each functional module domain. These operations are not performed sequentially as shown in the diagram; rather, all these adjustments are carried out simultaneously, depending on end-to-end stochastic gradient descent to optimize all of the model parameters in parallel in searching for an optimal solution.


The network shown in Figure~\ref{fig_003_mesoscale_model_convolutional_neural_network_architecture}.(b) consists of seven layers and represents only a part of the neural network architecture for learning a mesoscale model. The two layers labeled Roman Numeral 2 (II) and 5 (V) and shaded green encode the complete neural state vector at time $t$ and $t_{i + 1}$ respectively. Neural state vectors are derived from the functional data encoded in the sequence of transmission-response graphs that combine both structural and functional data. 

The bottom layer shaded in purple encodes a set of spatially-and-temporally-localized features computed from the current transmission-response graph that have proved to be effective in identifying functional motifs. These features assist in learning the functional domain boundaries. The layer labeled Roman Numeral 3 (III) and shaded orange is the {\it{configuration layer}} and is divided into sublayers shown here separated by dashed lines, each sublayer encoding the local parameters for one functional basis filter. 

Layer 4 (IV) is a convolution layer, shown here with three functional basis filters, each one associated with a unique functional module and its corresponding parametrically-configured neural network. The units in the penultimate layer labeled Roman Numeral 5 (VI) constitute the output layer of the model, divided into sublayers shown separated by dashed lines, one sublayer for each basis filter. The final layer (VII) scores the accuracy of each basis filter at each location. 

The basis filters are responsible for reconstructing the neural state vector of each recorded neuron. If there are 17 filters, each neuron is modeled by a sparsely-weighted linear combination of 17 filters so that only a few filters contribute significantly in accounting for any given neuron. Once trained, each filter applies the same globally-configured network everywhere, but employs a locally-configured interface network to adapt the globally-defined module to local conditions. 

Each fully-trained network represents a distinct functional motif, and the system as a whole can be deployed as a powerful motif finder. The framework allows us to explore a number of interesting hypotheses. By experimenting with the sparsity-inducing penalty term in the loss function, it should be possible to explore questions concerning whether all cortical columns compute the same function, or measure the diversity and distribution of function if it appears there is no single function. 


\subsection{Inferring Functional Module Domains}


\begin{figure}
  \begin{center}
    \includegraphics[width=\textwidth]{\figures/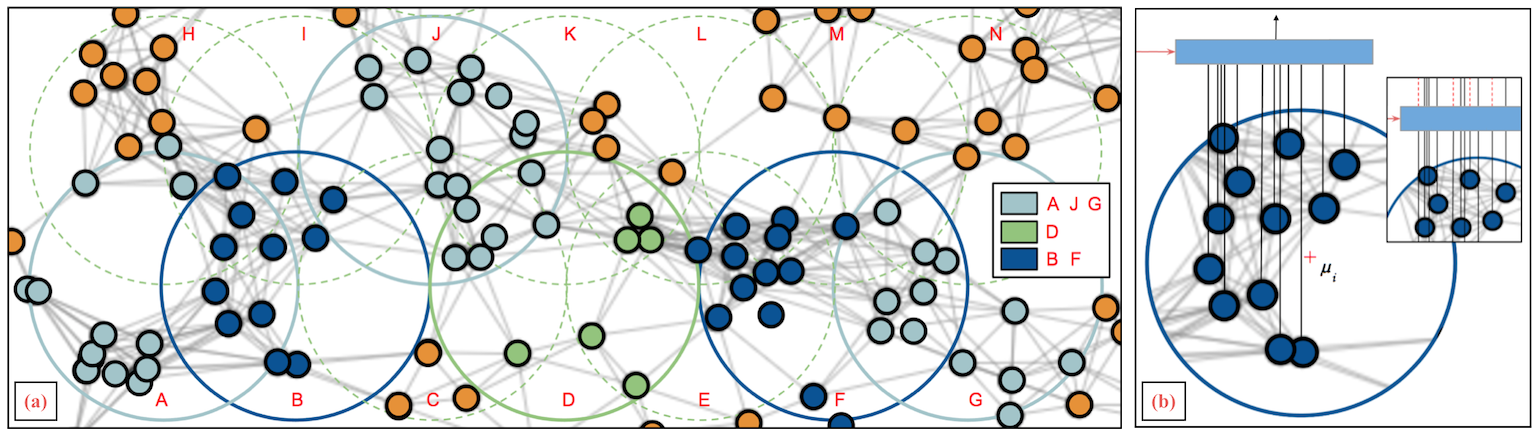}
  \end{center}
  \caption{The graphic on the left (a) illustrates how the functional basis filters decompose the connectome graph and associated point sources into functional domains. Each of the large dashed and solid circles represents a spherical subvolume of the connectome embedding space. When training is complete, each subvolume and each point source will be claimed by exactly one basis filter. In the graphic, three color-coded filters are shown claiming a total of six of the fourteen subvolumes. The stride of the sliding-window convolutional operator is half of the diameter of receptive-field subvolume. Note that with the exception of {\redhighlight{B}} none of the filters include\emdash{}and thus are responsible for modeling\emdash{}all the point sources in their subvolumes. The graphic on the right (b) illustrates an interface\emdash{}blue rectangle\emdash{}for a functional module corresponding to one or more recurrent layers that learn to assign domains for each location $\hmgp_i$ filter $f_j$ pair. The horizontal red arrow represents input from the configuration layer and vertical black arrow represents the output of local interface that determines the location-and-module-specific convolution matrix used to compute the filter / domain module output. Interfaces work by restricting the domain of $f_j$ when evaluated at $\hmgp_i$ to a subset of the cells that fall within the receptive field centered at $\hmgp_i$ where the dashed red lines in the inset depict cells that are excluded from the domain.}
  \label{fig_004_functional_module_domain_assignment_example_implementation}
\end{figure}


Training a mesoscale model involves learning a set of configuration parameters for each functional module network just as we learn the convolution matrix for each filter in a traditional filter bank. We also learn the sparse weights that define the linear combination of network outputs at each location in the input graph embedding, and, finally, we learn a separate functional module interface\emdash{}see Figure~\ref{fig_004_functional_module_domain_assignment_example_implementation}.(b)\emdash{}for every combination of module and receptive field in the 3D embedding of the transmission response graph.

We don't expect the fixed-diameter spherical subvolumes defined by the fixed 3D grid of coordinates to exactly partition the volume into functional domains. The filter diameter and stride that determine the grid of points are supplied by the researcher constructing the model and only serve as a proposal or upper bound on the extent of functional domains and by so doing constrain search. Each spherical subvolume encloses a unique maximal subgraph with a variable number of vertices and edges.

Since adjacent spherical subvolumes overlap, their enclosed subgraphs may share some vertices and edges. It likely that in some cases multiple functional modules will cover different portions, i.e., subsubgraphs, of the subgraph enclosed by any given spherical subvolume\emdash{}see Figure~\ref{fig_004_functional_module_domain_assignment_example_implementation}.(a). This complication highlights another issue relating to variability in the size and constitution of these maximally enclosed subgraphs.

The parameters of each configurable network instantiating a domain module are set globally. Each network is expected to represent a distinct functional motif and to contribute to predicting portions of the neural state vector in multiple, widely distributed locations throughout the target microcircuit. However, the enclosed subgraphs will vary and different regions of a microcircuit are likely to manifest different spiking frequencies, signal attenuation and noise characteristics.

The solution is to learn a local interface for each functional-module / enclosed-subgraph pair\footnote{%
  The local interface layers are not intended to perform any substantial computation otherwise they could subvert the primary reason for learning a sparse function basis, namely identifying repeated computational motifs that apply to broadly throughout the microcircuit. Interface layers serve to translate between local structural and signal characteristics and the global algorithmic characteristics of the functional modules that comprise the sparse function basis. To help enforce this restricted role the interface layers are designed to implement simple embeddings and are subject to a regularization term in the loss function.}.
The interface for a given module and enclosed subgraph aligns the inputs and outputs by selecting some neurons for inclusion in the module domain and rejecting others, modulates and shapes the local signal characteristics and, in analogy to matching a stereo amplifier to different inputs, e.g. microphones, musical instruments and recorded-music sources, and outputs, e.g., headphones, desktop speakers and home theater systems, matches impedance between the specified module and the selected inputs and outputs\footnote{%
  In electronics, impedance matching is the practice of designing the input impedance of an electrical load or the output impedance of its corresponding signal source to maximize the power transfer or minimize signal reflection from the load. Dynamic range is the range of acceptable or possible volumes of sound occurring in the course of a piece of music or a performance, and is generally measured as the ratio of the largest to the smallest intensity of sound that can be reliably transmitted or reproduced by a particular sound system, measured in decibels.}.


\subsection{Learning a Sparse Functional Basis}


\begin{figure}
  \begin{center}
    \includegraphics[width=\textwidth]{\figures/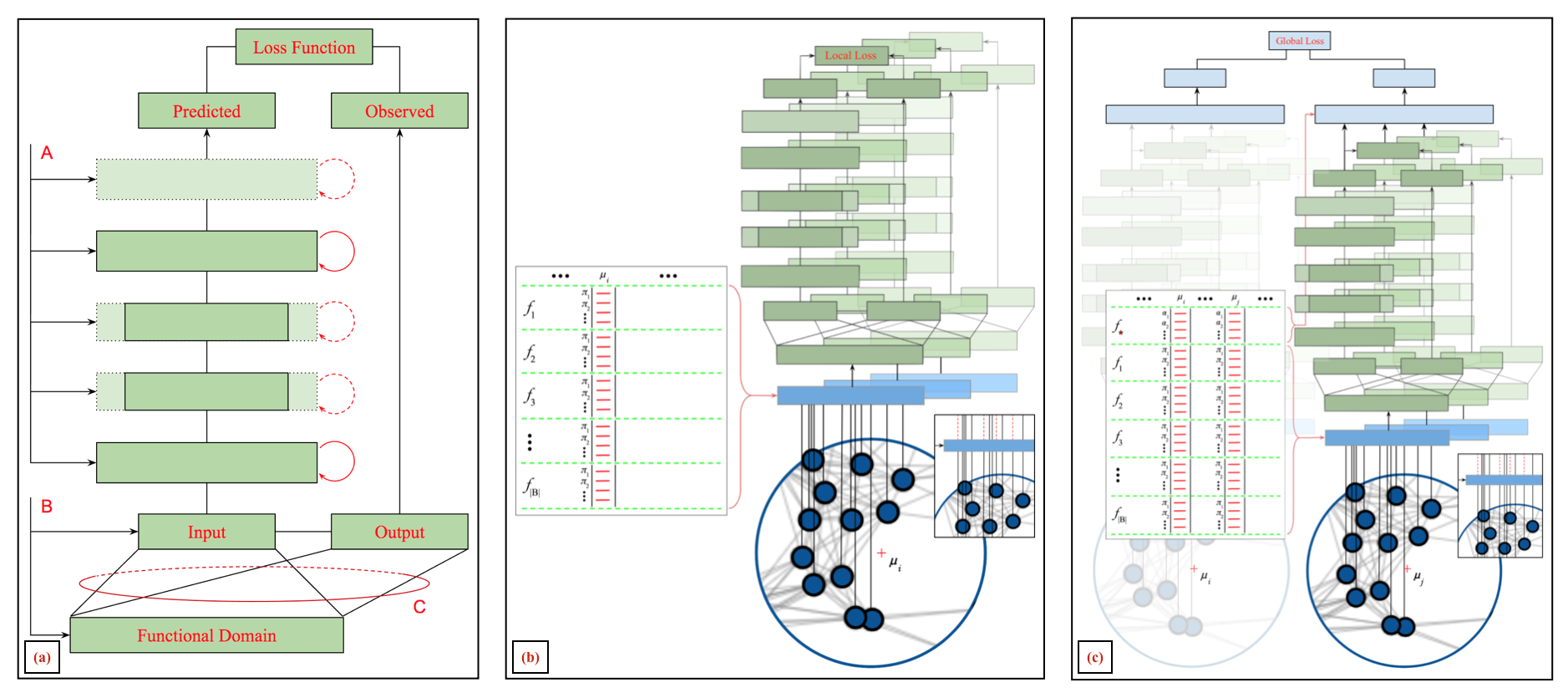}
  \end{center}
  \caption{Each filter in the sparse functional basis of a mesoscale model corresponds to a functional module implemented as a configurable network (a) that has one set of (global) parameters ({\redhighlight{A}}) that is the same for every location in the 3D grid spanning the connectome embedding and a second set of parameters ({\redhighlight{B}}) that defines location-specific properties. These parameters\emdash{}represented by the inset table on the left in panel (b)\emdash{} are learned along with all the other model parameters and reside in the configuration layer. The middle panel (b) represents the arrangement whereby the configurable networks corresponding to three different basis filters are scored on their ability predict the outputs from their inputs as defined by their respective interfaces. During training, these scores are reflected in the loss function and used to modify all the down stream adjustable parameters of the model via backpropagation. The right panel (c) illustrates how the same three basis filters operating on different locations compete to predict the same components of the neural state vector. The local and global loss are explanatory fictions as they are really just separate terms in a single loss function.}
  \label{fig_005_configurable_functional_modules_configuration_implementation}
\end{figure}


The configurable neural network of a functional module (filter) illustrated in Figure~\ref{fig_005_configurable_functional_modules_configuration_implementation}.(a) is defined by a set of learned global parameters ({\redhighlight{A}}) that apply to every location in the 3D grid spanning the connectome embedding and a second set of trained parameters ({\redhighlight{B}}) that determine location-specific properties described in Figure~\ref{fig_004_functional_module_domain_assignment_example_implementation}. The global parameters determine the number, size and type of layers using sigmoidal switches that change the number of units within a layer in fixed increments, eliminate layers altogether by enabling pass-through layers, add recurrent and skip forward edges, and select between half-wave rectification, divisive normalization, max pooling, logistic and other activation functions. Since there is only one fully configured network for any given filter at any particular time, we have added what we refer to as an impedance-matching embedding layer ({\redhighlight{C}}) specific to each location in the 3D grid that spans the connectome graph embedding space. This layer also accommodates variation in the size of the learned location-specific subgraph that defines the functional domain of the basis filter. 

The table of local and global parameters encoded in the configuration layer is illustrated in the inset on the left of panel (b) in Figure~\ref{fig_005_configurable_functional_modules_configuration_implementation}. Each functional module filter $f_i$ in the functional basis has a {\it{level}} in the configuration layer allocated to storing those parameters of the functional module that govern the local properties of the filter. Each location $\hmgp_j$ in the 3D grid of locations the identify the centroids of filter receptive fields corresponds to a {\it{column}} in the configuration layer and encodes configuration parameters specific to that location. Each location also includes static information concerning its associated maximally-enclosed subgraph and corresponding point sources. Each filter / location pair, $(f_i, \hmgp_j)$ stores a set of $K$ parameters $\{ \hmcp_k | 1 \hmleq{} k \hmleq{} K\}_{(i,j)}$ that determine the point sources assigned to the domain of the filter functional module. The local properties also include the parameters of the local impedance matching and I/O sorting network embedding layers. They don’t include global information about the number, size and type of layers that comprise the module network nor do they include the parameter values that define those layers. 

Panels (b) and (c) in Figure~\ref{fig_005_configurable_functional_modules_configuration_implementation} provide additional detail on how a sparse functional basis is trained. Three basis filters (configurable functional modules) are shown. Panel (b) focuses on how the three filters compete to predict the neural state vector and thereby apportion inputs and outputs. The predictions of the three filters are combined and scored by a local loss function \emdash{} more precisely an additive term in the global loss function, and gradients are propagated back through the network and used to adjust the configuration parameters including those that assign functional module domains. Panel (c) builds on (b) by illustrating the three filters applied to two locations labeled $\hmgp_i$ and $\hmgp_j$. The receptive fields at these two distinct locations may overlap, and, if so, the filters compete not just with one another but also with themselves at different locations. 

Ideally, when training is complete each point source is assigned to exactly one filter / location pair. To resolve conflicts, an additional (global) loss term\emdash{}shown in panel (c)\emdash{}is combined with the local loss to adjust domain assignments taking into account both a filter's performance in predicting its assigned portions of the neural state vector and its contribution to predicting the observed behavioral state. The configuration layer has been modified to include an additional layer labeled $f_{\rasterisk{}}$ assigns a scalar value in the unit interval to each basis filter that together determine a linear combination of the basis vectors responsible for predicting the behavioral state. Once again, propagating gradients back through the network serves to adjust the configuration parameters, and, in particular, adjust functional module domain assignments to encourage coverage and sparsity.


\begin{figure}
  \begin{center}
    \includegraphics[width=\textwidth]{\figures/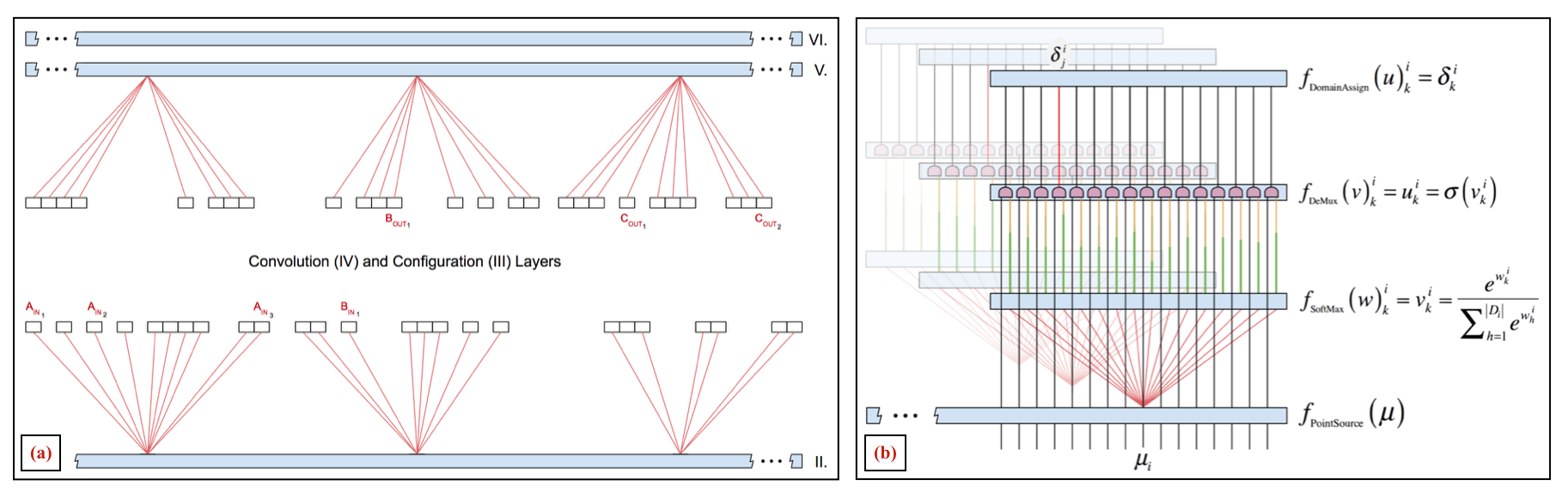}
  \end{center}
  \caption{The graphic on the left (a) provides architectural detail illustrating how one might model synaptic structure using two additional hidden layers: one layer is used to demultiplex the (observed) neural state vector at time $t$ represented in Level II onto a hidden layer representing the presynaptic domains of each neuron, and a second hidden layer to multiplex the postsynaptic domains onto the (predicted) neural state vector at $t + 1$ represented in Level V. The graphic on the right (b) suggests one way in which domain assignments might be accomplished using a short stack of layers corresponding to standard components, including a softmax layer and a gating layer that performs a function similar to the error carousel mechanism used in certain classes of recurrent neural networks. Technical detail concerning the latter is provided in Appendix~A.}
  \label{fig_006_synaptic_interface_layers_to_convolution_configuration_layers}
\end{figure}


There are many details concerning how to design and train a mesoscale model of the sort outlined in this paper. We have tried to address the most technically challenging problems by sketching plausible solutions based on our experience in building and deploying (deep) artificial recurrent neural networks or by building prototypes and experimenting with existing functional and structural datasets. 

Some of the data-related problems we summarize in Section~\ref{sec_discussion}, and, to conclude this section, we take a look at two questions that come up repeatedly in presenting this work in public forums: (a) if you can't record from synapses, is there any way you can still exploit the synaptic structure available in the connectome, and (b) how do you partition the connectome graph into sub graphs corresponding to functional domains without producing a combinatorial explosion.

\subsection{Representing Synaptic Structure}

Within the next few years, we should have technology allowing us to record changes in calcium concentrations at 30 FPS from most if not all of the neurons in an organism as complicated as a fly or larval zebrafish using genetically encoded calcium indicators expressed in the cell body nucleus or region around the axon hillock. Dense functional recordings of the activity in all of the synapses using calcium or voltage indicators at 30 FPS or higher will likely take a bit longer.

In the meantime, it would useful to exploit what we know about the synaptic connections between neurons. We will, after all, be able to infer a lot about the type and morphology of individual neurons, how the dendritic and axonal arbors are interconnected, and, while we can't resolve the activity in individual synapses, we can represent that latent activity in one or more hidden layers and attempt to infer the activity from a combination of structural and functional data.

Figure~\ref{fig_006_synaptic_interface_layers_to_convolution_configuration_layers}.(a) illustrates how one might model the structure of the synaptic network and infer latent activity relating to synaptic strength by employing two additional hidden layers. Both hidden layers have as many units, $S$, as there are synapses.

One hidden layer is inserted into the network shown in Figure~\ref{fig_003_mesoscale_model_convolutional_neural_network_architecture}.(b) directly below Layer~IV which has many units, $N$ as there are neurons / point sources. This hidden layer serves to mirror the pre-synaptic structure, so that the $N \times S$ hidden-layer weight matrix has an entry in the $i$th row and $j$th column if and only if the $i$th neuron is the pre-synaptic neuron for the $j$th synapse. 

The second layer is inserted into the network shown in Figure~\ref{fig_003_mesoscale_model_convolutional_neural_network_architecture}.(b) directly above Layer~II which as many units as there are neurons / point sources. This hidden layer serves to mirror the post-synaptic structure, so that the $S \times N$ hidden-layer weight matrix has an entry in the $i$th row and $j$th column if and only if the $j$th neuron is the post-synaptic neuron for the $i$th synapse. 

The first hidden layer leverages the synaptic structure to model the hidden state of individual synapses and thereby facilitate learning the sparse functional basis. The second hidden layer integrates predictions regarding synapses to update the neural state vector. The idea is relatively straightforward. We are looking for a dataset that includes both structural and functional recordings to evaluate the approach. 


\subsection{Synapses as Latent Point Sources}


\begin{figure}
  \begin{center}
    \includegraphics[width=3.75in]{\figures/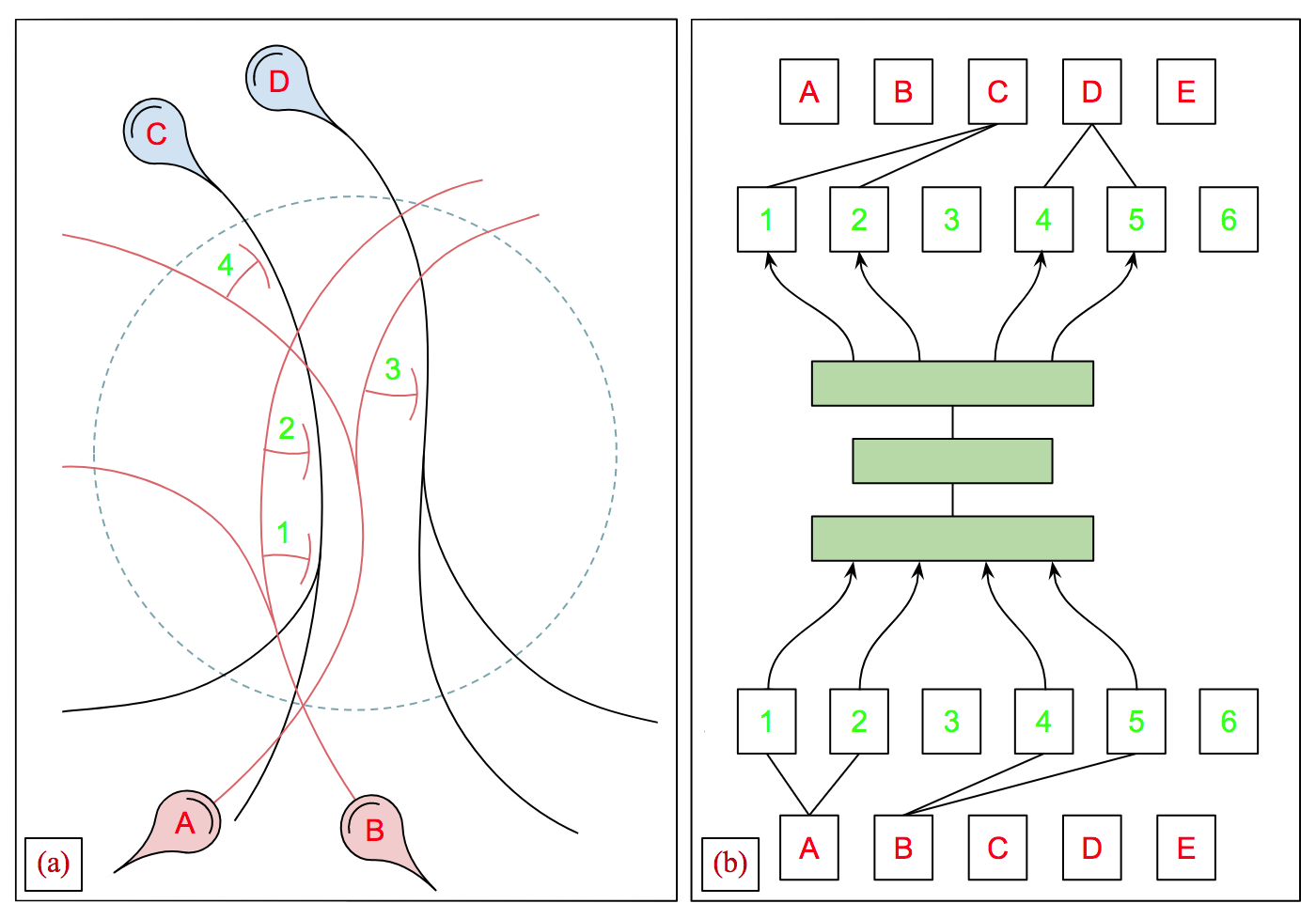}
  \end{center}
  \caption{The graphic on the left (a) is similar to Figure~\ref{fig_training_modules_microcircuits_architecture}.(b) with additional labeling to illustrate how we might exploit the synaptic information made available in the two hidden layers described in Figure~\ref{fig_006_synaptic_interface_layers_to_convolution_configuration_layers}.(a). The graphic on the right (b) illustrates how the units in these two hidden layers are mapped to individual synapses so they can be be utilized as latent point sources to define functional domains.}
  \label{fig_006_synaptic_interface_to_functional_module_domain_assignments}
\end{figure}


We have been using the term point source to refer to a location directly associated with neural activity, as in the case of a cell nucleus or synapse expressing fluorescent protein. We hypothesize that the activity and location of synapses offer important clues for inferring function, and that, by using our detailed knowledge of the local microcircuit provided by the structural connectome, we can relate activity in a neuron's cell body back to the locus of subsequent or prior activity in its synapses. Here we are propose that, given the arrangement of layers shown in Figure~\ref{fig_006_synaptic_interface_layers_to_convolution_configuration_layers}.(a), we can now treat {\it{inferred}} activity in synapses as point sources that constitute functional domains. 

Figure~\ref{fig_training_modules_microcircuits_architecture}.(b) is ambiguous regarding the semantics of the vertices shown in the microcircuit. The connectivity shown would suggest they are neurons, but the geometry is not clear and the obvious interpretation doesn't support our hypothesis. Here we complicate the picture somewhat by explicitly including synapses. Figure~\ref{fig_006_synaptic_interface_to_functional_module_domain_assignments}.(a) assumes the semantics of Figure~\ref{fig_001_structural_connectomics_and_microcircuit_connectome_graphs}.(c) with labels for neuron cell bodies (uppercase letters) and synapses (integer numbers). 

The dashed circle depicts a cross section of a spherical subvolume corresponding to a receptive field. Assume the four synapses shown within dashed circle are a subset of the set of all synapses enclosed in the spherical subvolume. Suppose further that these four synapses constitute the domain assigned a particular functional module. Specifically, the configurable network of this module applied to that receptive field is responsible for accounting for the computations performed by those four synapses illustrated schematically in Figure~\ref{fig_006_synaptic_interface_to_functional_module_domain_assignments}.(b).

This arrangement provides a good deal of flexibility in providing an interface to the functional recordings. The corresponding layers in Figure~\ref{fig_006_synaptic_interface_layers_to_convolution_configuration_layers}.(a) could provide a linear map or they could be fixed to reflect the information about the strength of connections, about the distance or expected time delay for signals traveling from the pre-synaptic cell body to the synapse or from the synapse to the post synaptic cell body. Generally it is a mistake to try to engineer neural network layers, e.g., by adding a layer of Gabor or Hermite functions in the case of modeling the early ventral visual stream. However, in this case, the intention is to simply supply the system with information provided by analysis of the structural data. 


\subsection{Domain Assignment Architecture}

In Figure~\ref{fig_005_configurable_functional_modules_configuration_implementation} we showed how functional modules are evaluated and assigned the responsibility for predicting different parts of the neural state vector. The top two layers of each stack of layers shown in green compute signals measuring the predictive accuracy of each functional module.

During back-propagation these signals are fed back and combined with additional local network features to assign cells to functional domains. These features are derived from the local properties of the static connectome graph and summary statistics of the functional time series characterizing the mutual information of adjacent subcircuits.

Figure~\ref{fig_006_synaptic_interface_layers_to_convolution_configuration_layers}.(b) illustrates how these signals are combined and used to route different point sources to different functional domains. The four (sub) layers shown in Figure~\ref{fig_006_synaptic_interface_layers_to_convolution_configuration_layers}.(b) are sandwiched between Layers II and III in the network shown in Figure~\ref{fig_003_mesoscale_model_convolutional_neural_network_architecture}.(b). 

Ideally, each point source is assigned to exactly one functional domain\emdash{}that is, one basis filter in one location in the 3D grid of locations. There are relatively few receptive fields that can contain any given point source. The number depends on the resolution of the 3D grid which depends on the size (diameter) and shape (spherical) of receptive fields and the step size (stride) of the sliding-window convolutional operator. For a given point source $\hmcp_{1}$, let's say there are $L_{1}$ possible locations and $B$ basis filters, then $\hmcp_{1}$ can be assigned to exactly one of $|D_{1}| = L_{1} \times{} B$ possible location-filter pairs.

The lowest sub layer in Figure~\ref{fig_006_synaptic_interface_layers_to_convolution_configuration_layers}.(b) is of size $N$, the number of point sources. The remaining three sub layers are of size $B \times{} \sum_{i}^{N} L_{i}$. Each set of possible location-filter pairs is handled by a separate softmax layer followed by a 1-of-$N$ gating layer to make the necessary assignments. 

Figure~\ref{fig_006_synaptic_interface_layers_to_convolution_configuration_layers}.(b) is more rough sketch than precise specification intended to match the level of detail afforded the other topics in this section. However, given the importance of the topic for the success of the proposed project, we highlight the subject of domain assignment in Appendix~A providing additional technical detail and a discussion of the algorithmic alternatives.


\subsection{Tacit  Independence Assumptions}

Supplement~B outlines the major factors we believe will determine the success or failure of a project organized around the ideas presented in this paper. The architectural features illustrated in Figure~\ref{fig_006_synaptic_interface_layers_to_convolution_configuration_layers} raise some specific technical challenges that warrant discussion here while the concepts introduced in Figure~\ref{fig_006_synaptic_interface_layers_to_convolution_configuration_layers} are still fresh. These specific challenges pertain to functional / relational independence of one sort or another. We are unlikely to find strict functional dependence between any two parts of the brain. Our goal here is to make it easier to think about and possibly quantify the consequences of functional models that violate our tacit independence assumptions.

The domain assignment architecture illustrated in Figure~\ref{fig_006_synaptic_interface_layers_to_convolution_configuration_layers}.(b) operates on the connectome graph $G = (V,E)$ in which the vertices, $V$, correspond to neurons and the edges, $E$, correspond to synapses. The domain-assignment network attempts to partition / decompose $G$ into subgraphs that define the domains of functional modules that implement the basis filters. Decomposition requires assigning each neuron to a unique location-filter pair, implicitly assuming that the inter-module dependencies are completely determined by the inputs and outputs of the modules that are responsible for learning functions of different subgraphs. This may prove hard to achieve in practice.

Figure~\ref{fig_006_synaptic_interface_layers_to_convolution_configuration_layers}.(a) might seem to suggest a solution. The introduction of the two hidden sub layers exploits our knowledge of neural structure from the connectome to tie synapses that we cannot record from to neurons that we can. The weights that determine the two hidden layers map pre-synaptic neurons in the bottom hidden layer to their respective outgoing synapses and post-synaptic neurons in the top hidden layer to their respective incoming neurons\footnote{%
  Imagine two clothes lines running parallel to one another. In the center between the two clothes lines lying on the grass is a brain that has been expanded a thousand-fold so that each neuronal process\emdash{}axon or dendrite\emdash{}is the diameter of a fine thread. For each synapse, surgically separate the pre-synaptic side of the synapse from the post-synaptic side and prepare two lengths of string, each one half as long as the distance between the two clothes lines. Take the first string, sew one end to the pre-synaptic side of the synapse and tie the other end to the clothes line on the left. Take the second string, sew one end to the post-synaptic side and tie the other end to the clothes line on the right. Continue until you've carried out the same steps for every synapse in the brain. The clothes line on the left (right) corresponds to the bottom (top) hidden layer in Figure~\ref{fig_006_synaptic_interface_layers_to_convolution_configuration_layers}.(a).}.
As such, the corresponding (sparse) weight matrices could implement (bottom) simple individually-weighted incoming contributions or (top) composite cumulatively-weighted outgoing contributions.

If we were to allow a more densely-connected hidden layer, say, a bipartite-graph connection matrix, it might be argued on the basis of the universal approximation theorem for artificial neural networks~\cite{CybenkoMCSS-89,HorniketalNEURAL-NETWORKS-89} that the two hidden layers combined together could be trained to reproduce the transfer function for the entire microcircuit. As it is, the hidden layers are relatively simple and could potentially be made even simpler by setting all the weights to one\footnote{%
  The two hidden layers constitute, respectively, a {\it{demultiplexer}} that separates the contributions {\it{from}} a given neuron into a set of outputs corresponding to the pre-synaptic sides of its incident synapses, and a {\it{multiplexer}} that combines the contributions {\it{to}} a given neuron into a set of inputs corresponding to the post-synaptic sides of its incident synapses.}.
What if we defined the domain (inputs) and range (outputs) of functional modules in terms of synapses instead of neurons, and used an allocation strategy analogous that described in Figure~\ref{fig_006_synaptic_interface_layers_to_convolution_configuration_layers} for making domain assignments? What sort of independence relationships would such a network exhibit? 

Globally defined functional modules implementing basis functions require a local interface for every location in which they are applied. This interface adapts the local conditions to deal with different subgraphs and modulates and shapes the local signal characteristics. Configuring the neural network architecture for such interfaces has to be done with some care since otherwise the interfaces could be trained do all the work, i.e., each local interface could end up computing its corresponding local function independently and the corresponding basis function end up as the identity function. We can avoid this sort of degenerate solution, by limiting the capability of the general interface network architecture in much the same way that we discussed limiting the hidden layers in the synaptic interface shown in Figure~\ref{fig_006_synaptic_interface_layers_to_convolution_configuration_layers}.(a).

But what are we really expecting of these interface networks? The inputs are defined by the domain of the corresponding functional module. There are as many individual interfaces as there are grid-location / basis-function pairs. 
Whatever architecture we choose it will have to handle domains corresponding to different size sub graphs, varying numerical range, etc. It may be that all we need is a linear transformation from $\R^{N}$ to $\R^{M}$ where $N$ is the size of the domain and $M$ is the size of the input layer of the configurable network for the corresponding functional module \emdash{} recall that both $N$ and $M$ will change during training as the domain and network evolve, followed by a suitable squashing function or alternative strategy to control the numerical range. A one-size-fits-all decorrelaton and dimensionality reduction such as the Karhunen-Lo\`{e}ve transform~\cite{SirovichandKirbyJOSA-87} might suffice\footnote{%
  The reader might be concerned that the number of model parameters will be larger than the number of target features. However, if the model is doing its job of identifying and exploiting frequently occurring computational motifs, then it is more likely the target features will substantially outnumber the model parameters. Suppose we have a model with $100$ basis filters, each one having $10,000$ parameters. That amounts to a total of $1,000,000$ (1M) parameters not counting the parameters defining the interface layers, assuming for sake of argument that the interfaces are implemented as a one-size-fits-all local dimensionality reduction and decorrelaton transform adding a constant number of parameters independent of the size of the basis. If we assume a target with 1M neurons each with 1K synapses, then there is a 1000:1 ratio of target features to model parameters. If the basis filters are general enough then the ratio will only widen as the target increases in size.}. 
Would that it were so simple?\footnote{%
  Josh Brolin as the "fixer", Eddie Mannix, teaching Alden Ehrenreich as the "silent screen cowboy" turned Hollywood movie star, Hobie Doyle, the intricacies of the English subjunctive case in the Coen Brothers' {\it{Hail Caesar!}}}





\label{sec_discussion}
\section{Discussion}    


The model described here represents a theoretical exercise in how we might build machines to automatically infer the function of biological networks in much the same way that John von Neumann’s universal replicator~\cite{Neumann1966} provided a theoretical framework to explore questions concerning self-replicating machines. Von Neumann made use of the rather modest set of computational abstractions available at the time, and, when necessary, developed new tools including cellular automata which he co-discovered with Stanislaw Ullam.

In the case of exploring the possibility of inferring mesoscale models, we have access to a powerful set of computational tools covering a wide range of abstractions in order tackle what is arguably a more complicated computational process than self replication. Specifically, advances in the theory and application of artificial neural networks have made it possible to use the corresponding abstractions and implementations to blur the line between the two, simplifying implementation and modeling while at the same time offering mesoscale abstractions that can be computationally manipulated and analyzed. 

In von Neumann's equally-influential work on synthesizing reliable organisms from unreliable components~\cite{Neumann1956} he sought to draw inspiration from biology but considerably simplified his analysis by using the model of neural activity developed by McCulloch and Pitts~\cite{McCullochandPitts43}. While von Neumann's analysis and McCulloch and Pitt's neuron model have perhaps influenced the field of neuroscience less than computer science~\cite{PippengerAMS-90}, it was von Neumann's insight about computing with unreliable components and Carver Mead's related observations on neuromorphic computing~\cite{MeadIEEE-90,MeadCIT-87} that inspired many of the simplifications and abstractions implicit in the account of neural computation featured in this paper.

Implementations of the ideas described here will depend on the development of advanced technology to record enough of the right sort of data to discern statistical patterns that will allow us to to infer computational abstractions of some value in bridging the gap between the micro and macroscale. Success will be due in large part to the concessions we make in accommodating the diversity and complexity of the biological processes we seek to explain. It is certainly possible that the class of mesoscale models presented here have required concessions that would be deemed unacceptable for other use cases. There is no one single class of mesoscale models but rather a diverse collection spanning a wide spectrum of abstractions.

Our primary objective is to build tools that, given structural, functional and behavioral recordings obtained from a biological organism, can infer a model of the target system that can be used to simulate the system, predict its behavior from sample input, and decompose said system into a collection of components that are pure functions\footnote{%
  A pure function always evaluates to the same output given the same input. Strictly speaking a pure function can't depend on any hidden state that changes during evaluation or between separate executions. Since most neural computations are stateful and recurrent, we side-step the strict definition by assuming the recurrent state is always provided as one of the inputs.}
of local inputs and outputs, that taken together fully determine the dependencies among components. Using such tools, scientists will be able to construct models by creating datasets, developing abstractions that correspond to families of functions realized as configurable neural networks, and adjusting hyper parameters that constrain the domain and range of those function to test hypotheses and gain insight into the emergent properties arising from the interactions among individual neurons.

Why you might ask have we gone to the trouble to provide all this detail. Why didn't we simply finish writing the code, run the experiment, and present the result. For one thing, this is a theoretical paper. Its primary goal is to present a class of models and describe how one might go about inferring such models from functional and structural data. We are quite confident that we are not the only researchers able to conceive of such a model and part of the reason we present our model in its present state is because we want to start a conversation about what sort of models are really needed and why one would want to build one in the first place.

Another reason why we didn't complete the model is that it will likely take several years before we get to the point where we have the data and the necessary infrastructure to learn a mesoscale model of anything as complex as a fly. We have been working on connectomics for over three years now and, while we have made enormous progress and hope to publish a significant fraction of the connectome of the fly working with our collaborators at HHMI Janelia Research Campus within the next year, we now have a great deal of experience building technology capable of training and testing very large-scale neural networks and developing related stitching and alignment software of the sort necessary to perform fully automated reconstruction of all the neuropil in a fly brain. Needless to say, we have a great deal of respect for the problems we propose to solve in this paper.

We could start out with simpler organisms or selected small tissue samples of well understood function. We have done smaller experiments trying to understand the challenges of solving our more ambitious goals, and we will continue to do so even as we seek to mount a larger effort directed at solving problems of the scope described in this paper. We seek to solve the larger problem because we believe we can and see little sense in our tackling problems that other labs could easily take on. Part of the rationale for mounting projects in basic science at Google is to push our infrastructure and technology to its limits and beyond, for we believe that those limitations will arrive sooner than later. Working on ambitious projects also allows us to partner with scientists at other institutions who share our aspirations and recognize the advantages of collaborating with industrial partners who are used to tackling complex software engineering problems and working with very large datasets. Supplement~B provides a more detailed account of the challenges we believe will determine the success or failure of this project.



\label{sec_conclusion}
\section{Conclusion}


The framework described in these pages provides a powerful tool for conducting exploratory and hypothesis-driven science relating to complex dynamical systems, not just in the field of neuroscience. It can be used to discover emergent computational strategies by exploiting levels of abstraction afforded by different modeling languages. It is designed to identify computational motifs in the form of general computational strategies that roughly correspond to conventional algorithms. These motifs are realized in statistically similar structural and functional patterns that manifest throughout the target, often arranged in a regular lattice such that each element of the lattice is similar but not identical and implements functionally similar but not identical computational capabilities. 

We expect both structural and functional recordings will provide clues regarding computational similarity despite variability due to the stochastic processes by which they were constructed during early development and shaped through interaction with the external environment. We believe it is possible to induce computational accounts that provide useful explanations of observable behavior and abstract from the variability we observe at the microscale. Variants of the tools described here could be used to understand changes in migratory patterns as a function of changes in climate or habitat or the spread of infection taking into account microbial adaptation, examples in which computational processes play out on multiple spatial and temporal scales and are highly influenced by the rate at which information is transmitted and shared between the entities that comprise the underlying system.



\label{sec_acknowledgments}
\addcontentsline{toc}{section}{Acknowledgments}
\subsection*{Acknowledgments}

Thanks to all those who hosted early presentations of this work at conferences they organized or at their home institutions, including
Kristofer Bouchard,
Matt Botvinick,
Ed Boyden,
Mark Churchland,
David Cox,
Ken Harris,
Michael Hausser,
Adam Marblestone,
Bill Newsome,
Bruno Olshausen,
Olaf Sporns,
Gerry Rubin, and Sebastian Seung.
While some parts of the model described here have yet to be implemented, several of the component pieces were prototyped using datasets, tools and insights from the following researchers,
Misha Ahrens,
Sophie Aimon,
Michael Buice,
Vivek Jayaraman,
Saul Kato,
Jeff Lichtman,
Marius Pachitariu,
Michael Reiser,
Stephen Plaza,
Shinya Takemura and Manuel Zimmer.
Thanks to all of those who commented on early versions of this work, asking insightful questions and contributing useful ideas and references, they include 
Shaul Druckmann, 
Michal Januskewski,
Peter Latham,
Wei-Chung Lee,
Liam Paninski,
Art Pope,
Jon Shlens,
Fritz Sommer,
Karel Svoboda,
Maneesh Sahani,
Rahul Sukthankar,
Dan Yamins and Jay Yagnik.





\label{sup_point_source_domain_assignment_networks}
\addcontentsline{toc}{section}{Supplement A: Automated Domain Assignment}    
\section*{Supplement A: Automated Domain Assignment}    


This supplement provides additional detail concerning how point sources corresponding to cell bodies identified in the recorded functional are assigned to the domains of functional modules operating on specific receptive-field locations. This aspect of mesoscale modeling was first addressed in Figure~\ref{fig_003_mesoscale_model_convolutional_neural_network_architecture}.(b) and is supplemented here by providing a more detailed description of the network architecture and including a complementary mathematical characterization defining the functions computed in each layer.

In the following, we use capitalized "Layer" to refer to the Roman-numeral-numbered layers in Figure~\ref{fig_003_mesoscale_model_convolutional_neural_network_architecture}.(b) and lower-case "layer" to refer to what are essentially sub layers or detail-oriented layers between the higher level Layers. The top layer (a) is recurrent and serves to approximate the neural state vector at time $t$ from the functional input (recorded point sources) at time $t$ and the neural state vector from $t-1$. The penultimate layer (b) is a bank of differentiable gates that route input (individual point-sources) to location-specific applications of particular basis filters (functional modules) identified in terms of grid-location / basis-filter pairs.

Indeed, neither the inputs to nor outputs from layers (a) and (b) correspond to adjustable (trainable) weights in our model. We could eliminate them altogether by packing more complexity into the activation function of layer (c). In the expanded functionally-separated view, layer (c) implements multiple softmax functions, where each softmax function in combination with a subset of the gates in layer (b) assigns each point source to exactly one grid-location / basis-filter pair thereby specifying the domains for all functional module (instances) applied in performing the updates at time $t$\footnote{%
  The softmax functions in layer (c) together with the loss function and the differentiable gates in layer (b) implement a sparse functional basis where the degree of sparseness depends on loss function and optimization method used in training. To simplify the discussion here, we assume that the resulting predictions in Layer VI are computed at each grid location as a weighted sum of the output of each filter where exactly one weight is one and the rest zero. The consequence of relaxing this assumption is that a less punctate weighted sum is more difficult to interpret and arguably less useful as part of an explanatory theory.}.
Technically, (a), (b) and (c) could be implemented as a single recurrent hidden layer implemented as a Long Short-term Memory (LSTM) model using LSTM constant error carousels as the gates in layer (b)~\cite{SchmidhuberDNN-14,HochreiterandSchmidhuberNC-97,HochreiterPhD-91}. 

Layers (d) and (e) are more conventional neural network layers with adjustable weights that we learn by backpropagation. These layers are not, however, as simple as the more common bipartite-graph or convolutional layer. Prior to discussing layers (d) and (e) it will help if the reader is familiar with the terms introduced in the main text and has a better understanding of how they relate to one another. We employ the following notational conventions where each element of the model relevant to the discussion at hand is listed in the format: individual-element symbol, related set-of-elements symbol if relevant $\hmrarr{}$ short-form reference in italics, long-form acronym in brackets, and a brief explanation: 
\begin{itemize}
\item $\hmps{}$, $\hmPs{}$ $\hmrarr{}$ {\it{point source}}, [locus of recorded activity] [LCA] $\hmrarr{}$ Simplifying, each point source corresponds to the coordinates of a cell body nucleus identified in the functional data by its signature fluorescent-emission spectra.
\item $\hmgp{}$, $\hmGp{}$ $\hmrarr{}$ {\it{grid location}}, [receptive-field centroid] [RFC] $\hmrarr{}$ Convolution requires a 3D grid of locations defined by the dimensions of the sample, diameter of a spherical receptive field and stride of the sliding-window operator.
\item $\hmbf{}$, $\hmBf{}$ $\hmrarr{}$ {\it{basis filter}}, [functional module basis] [FMB] $\hmrarr{}$ Each model includes a functional basis consisting of functional modules, defined by a domain and range of point sources, an interface, and a configurable neural network.
\item $\hmcp{}$, $\hmCp{}$ $\hmrarr{}$ {\it{configuration parameter}}, [module configuration parameter] [FMC] $\hmrarr{}$ Each basis filter is a functional module whose component domain, range, interface and network are defined by a collection of learned configuration parameters.
\item $\hmgbp{}$, $\hmGbp{}$ $\hmrarr{}$ {\it{grid-basis pair}}, [grid-location / basis-filter pair], $\hmgbp{}_{(i, j)} = (\hmgp_i, \hmbf_j)$ [GBP] $\hmrarr{}$ Each point source is assigned to the domain of one functional-module basis filter operating on the receptive field containing the point source centered at one grid location.
\item $\hmdgf{}$, $\hmDgf{}$ $\hmrarr{}$ {\it{gather network}}, [domain-assignment gather function] [DGF] $\hmrarr{}$ The neural network responsible for assigning grid-location / basis-filter pairs to a given point source collects information from a relatively small set of nearby point sources.
\item $\hmdsf{}$, $\hmDsf{}$ $\hmrarr{}$ {\it{scatter network}}, [domain-assignment scatter function] [DSF] $\hmrarr{}$ The neural network responsible for assigning grid-location / basis-filter pairs to a given point source distributes information to each possible grid-location / basis-filter targets.
\item $\hmdcf{}$, $\hmDcf{}$ $\hmrarr{}$ {\it{composite network}}, [domain-assignment composite function] [DCF] $\hmrarr{}$ Alternative having a separate neural network responsible for assigning grid-location / basis-filter pairs to a given point integrating scatter and gather in a single composite network.
\end{itemize}

As an estimate of layer size and exercise in familiarizing the reader with the notation, the number of inputs to layers (a) and (c) correspond to the number of possible individual point-source domain assignments and is bounded by $max_{\hmgp{}} | \hmPs{}_{\hmgp{}} | \hmtimes{} | \hmGp{} | \hmtimes{} | \hmBf{} |$ where $\hmPs{}_{\hmgp{}}$ is set of point sources enclosed in the spherical receptive field centered at grid location $\hmgp{}$. The number of inputs to layer (b) is twice the number of inputs to layer (a). The number of inputs to each of the layers (d) and (e) is the same as the number of point sources or $| \hmPs{} |$. 


\subsection*{Assigning Point Sources}


\begin{figure}
  \begin{center}
    \includegraphics[width=\textwidth]{\figures/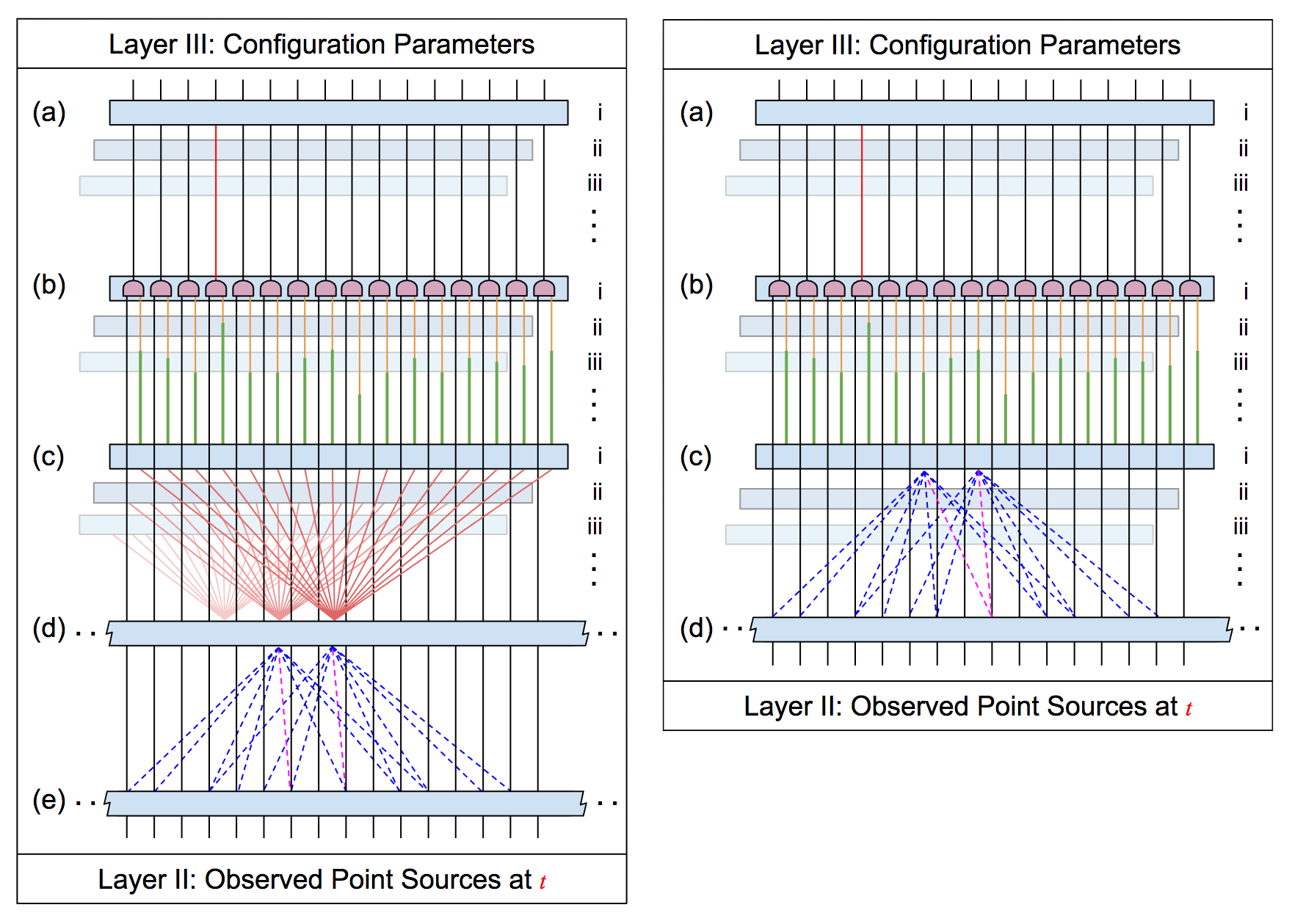}
  \end{center}
  \caption{This figure illustrates two alternative methods for automatically assigning point-sources\emdash{}such as the cell bodies of neurons identified in the functional recordings\emdash{}to functional module domains by adding sub layers (a), (b), (c), (d) and, in the case of the alternative shown in left panel, layer (e) sandwiched between layers II and III as described earlier in Figure~\ref{fig_003_mesoscale_model_convolutional_neural_network_architecture}.(b). The alternative on the left uses fewer adjustable weights, but may confer an advantage by sharing weights within and across different domains, while the alternative on the right may perform better if the functionally-relevant features vary substantially among different functional modules and locations within the tissue where their corresponding configurable neural networks are applied.}
  \label{fig_007_point_source_module_assignment_sublayers_between_layers_II_III}
\end{figure}


The layers (a), (b) and (c) are shown in Figure~\ref{fig_007_point_source_module_assignment_sublayers_between_layers_II_III} as partitioned into blocks numbered i, ii, iii, etc. In each of layers (a), (b) and (c), there is one such block for each point source of size $| \hmGbp{}_{\hmps{}} |$, where $\hmGbp{}_{\hmps{}}$ is the set of all grid-basis pairs associated with point source $\hmPs{}$. The layer (c) blocks implement the softmax function and each layer (c) block has its corresponding layer (b) block that implements a bank of gates that make the final selection.

The model on the right combines the functions of the weights connecting layers (e) and (d) and those connecting (d) and (c) in the model on the left, adding weights so that there is a dedicated (sub) network for each combination of a point-source $\hmps{}$ and one of its related grid-basis pairs $\hmgbp{} \hmisin{} \hmGbp{}_{\hmps{}}$, rather than a single shared network for each point source as in the model on the left.

In the connections between layers (d) and (c) there is exactly one adjustable weight connecting each point source to each of its possible grid-basis pair possibilities, and so the weights connecting (d) and (c) implement the following {\it{scatter}}\footnote{%
  Gather-scatter is a type of memory addressing that often arises when addressing vectors in sparse linear algebra operations. It is the vector-equivalent of register indirect addressing, with gather involving indexed reads and scatter indexed writes. ({\urlh{https://en.wikipedia.org/wiki/Gather-scatter_(vector_addressing)}{SOURCE}})} function(s):
\begin{equation}\label{eqn_scatter_function}
  \hmforall{} \hmps{}_{i} \hmisin{} \hmPs{},  
  \hmforall{} \hmgbp{}_{k} \hmisin{} \hmGbp{}_{\hmps{}_{i}},
  \hmdsf{}(\hmdgf(\hmps{}_{i}),\hmgbp{}_{k}) = \hmdgf(\hmgbp{}_{i}) \hmtimes{} w_{(\hmps{}_{i},\hmgbp{}_{k})}
\end{equation}

Layers (a), (b) and (c) have $\sum_{\hmphi{} \hmisin{} \hmPhi{}} | \hmGbp{}_{\hmphi{}} |$ units each, and layers (e) and (d) $| \hmPhi{} |$ each. The set of grid-basis pairs for a given point source is denoted $\hmGbp{}_{\hmps{}_{i}}$ and enumerated as $\{ \hmps{}_{(i,1)}, \hmps{}_{(i,2)}, \ldots{}, \hmps{}_{(i,|\hmGbp{}_{\hmps{}_{i}}|)} \}$. Note it is possible that $|\hmGbp{}_{\hmps{}_{i}}| \hmneq{} |\hmGbp{}_{\hmps{}_{i \neq{} j}}|$ and hence it is generally not possible to arrange the units comprising such a layer as a simple matrix or tensor without padding. When necessary, we index the units by function and using indexed arguments: layer (a) as $\hmlang{} \hmps{}_1, \hmps{}_2 \ldots{}, \hmps{}_{|\hmPs{}|} \hmrang{}$, layer (d) as $\hmlang{} \hmdgf{}(\hmps{}_1), \hmdgf{}(\hmps{}_2), \ldots{}, \hmdgf{}(\hmps{}_{|\hmPs{}|}) \hmrang{}$ and layer (e) as $\hmlang{} \hmdsf{}(\hmdgf{}(\hmps{}_1), \hmgbp{}_1), \hmdsf{}(\hmdgf{}(\hmps{}_1),\hmgbp{}_2), \ldots{}, \hmdsf{}(\hmdgf{}(\hmps{}_1),\hmgbp{}_{|\hmPs{}|}) \hmrang{}$.

The weights connecting layers (e) and (d) combine additional information from nearby point sources to implement the following {\it{gather}} functions, one for each point source:
\begin{equation}\label{eqn_gather_function}
  \hmforall{} \hmps{}_{i} \hmisin{} \hmPs{},  
  \hmdgf{}(\hmps{}_{i}) = w_{(\hmps{}_{i},\hmps{}_{i})} \hmtimes{} \hmps{}^t_{i} +
  \hmsum_{\hmps{}_{j} \hmisin{} \hmDgf{}_{\hmps{}_{i}}} w_{(\hmps{}_{i},\hmps{}_{j})} \hmtimes{} \hmps{}^t_{j}
\end{equation}
where $\hmps{}^{t}$ is the value of $\hmps{}$ at time $t$.

In the approach with five layers on the left in Figure~\ref{fig_007_point_source_module_assignment_sublayers_between_layers_II_III}, each point source has its own specialized context that may include, for example, all of the neighboring point sources within a radius of $10 \hmmicron$. In the approach with four layers, each combination of a point-source plus one of its grid-basis pairs has its own context trained to account specifically for the relative location of the point source with respect to the spherical receptive field centered at the grid-location coordinates. To implement the four-layer approach, we combine the scatter-gather functions into one composite layer:
\begin{equation}\label{eqn_composite_function}
  \hmforall{} \hmps{}_{j} \hmisin{} \hmPs{}, 
  \hmforall{} \hmgbp{}_{k} \hmisin{} \hmGbp{}_{\hmps{}_{i}},
  \hmdcf{}(\hmps{}_{i}) = w_{(\hmps{}_{i},\hmps{}_{i},\hmgbp{}_{k})} \hmtimes{} \hmps{}^t_{i} +
  \hmsum_{\hmps{}_{j} \hmisin{} \hmDgf{}_{\hmps{}_{i}}} w_{(\hmps{}_{i},\hmps{}_{j},\hmgbp{}_{k})} \hmtimes{} \hmps{}^t_{j}
\end{equation}
but not without using more parameters per point source, since we now have a separate (sub) network dedicated to each combination of a point-source plus one of its grid-basis pairs.

Using fewer adjustable weights as in the five-layer approach may confer an advantage by sharing weights within and across different domains, while the alternative four-layer approach may perform better if the functionally-relevant features vary substantially among different functional modules and locations within the tissue where their corresponding configurable neural networks are applied.


There are also options for enriching the set of features we use to represent graphs. Referring to Level~I in Figure~\ref{fig_003_mesoscale_model_convolutional_neural_network_architecture}.(b) labeled "Topological Invariant Features", we have explored several strategies for improving either the four- or the five-layer approach by incorporating (i) precomputed localized topological invariant features of the static connectome, e.g., see~{\urlh{http://web.stanford.edu/class/cs379c/resources/neurobot/}{here}} {\cite{DlotkoetalCoRR-16,SizemoreetalCoRR-16,CangandWeiCoRR-17}}, or (ii) learned features of the dynamic functionally-relevant transmission-response graph using graph convolutional networks, e.g., see~{\urlh{https://tkipf.github.io/graph-convolutional-networks/}{GCN}} {\cite{KipfandWellingICLR-17,DefferardetalNIPS-16}}.


\subsection*{Algorithmic Alternatives}


In general, our response to overcoming the computational challenges faced in building mesoscale models amounts to pointing out that we are already building models whose size, in terms of the number of parameters and amount of data required for training, rivals that of the models we propose here. We acknowledge that traditional measures of tractability based upon asymptotic worst-case runtimes are not the standard by which we should judge the anticipated computational effort. We routinely solve instances of problems that are classified NP-hard if not more challenging. Conversely we routinely solve problems classified as $O(n \log{} n)$ that are challenging simply because $n$ is so large.

Most of the algorithms that we apply correspond to approximations and their runtimes are generally determined heuristically. Software engineers in industry often have a pretty good idea what will work to satisfy a given measure of precision, recall and latency. Our experience at Google is that for problems of the type and size we are considering here\emdash{}targeting organisms with less than a million neurons and fewer than a billion synapses, we are reasonably confident that good software engineering will find a way to generate solutions that are good enough to suit our purposes. 

That said, there are some unknowns that lead us to be somewhat cautious, if only to allocate sufficient development time to engineer tractable solutions. Learning mesoscale models requires our approximating some particularly difficult optimization problems. As an example, the problem of learning a sparse functional basis is generally characterized as a non-convex optimization problem, and algorithms for solving such problems employ approximations that are well enough characterized that scientists and engineers employ them routinely. 

For example, alternating minimization (AM) is a general method used for optimizing two variables jointly that is often applied to sparse coding problems. Given two variables $p$ and $q$, the method works by alternating between solving for $p$ and $q$; more precisely, keep $p$ constant and optimize for $q$, keep $q$ constant and optimize for $p$, and continue alternating until the algorithm converges. Expectation maximization (EM) is a special case of AM ~\cite{TusnadyandCsiszarSDSI-84}, and variants of AM are among the most efficient for solving this class of problems~\cite{AroraetalPMLR-15}.

In the case of sparse coding, $p$ might correspond to the weights of the linear combinations of basis functions used to reconstruct the data and $q$ to the configurable network parameters that instantiate the functional modules comprising the basis. If this were all there was to the problem, we could probably use some variant of AM to solve it~\cite{BaoetalPAMI-16,LeeetalNIPS-06}. However, in our case, the problem of learning a sparse basis is complicated by the fact that we also have to learn the functional module domains\footnote{%
  Throughout the document we've stretched the term "domain" to encompass both the domain and range of a function, its inputs and outputs. The elements of the domain and range are point sources corresponding to vertices in the connectome graph. Taken together, they define a minimal (cut) set of edges such that, if you remove those edges from the connectome graph, you obtain a disconnected sub graph containing the domain and range. In general, this sub graph contains not only the corresponding domain and range vertices, but all the remaining vertices reachable from the domain and range vertices / point sources in the isolated sub graph.}.

Fortunately, we don't have to search the combinatorial space of all possible domains, since we use a {\it{restriction bias}} to limit search. Specifically, we start with the set of all spherical volumes corresponding to a receptive field centered at a 3D point in the convolution grid. For each such volume, we consider the set of proposals corresponding the set of all possible completely enclosed sub graphs. The resulting set of all proposals for each point in the convolution grid is still large, but significantly smaller than the set of all possible proposals without the restriction bias.

The domain assignment problem is at least as hard as the {\it{graph partition}} problem in theoretical computer science which is known to be NP-hard\cite{FederetalSTOC-99,GareyandJohnson79}. That's not surprising, nor is it particularly discouraging. We are assuming that there is a certain amount of redundancy in the neural microcircuit and hence there will likely be multiple solutions and we are not committed to finding the optimal solution. 

The problem of decomposing a microcircuit into functional module-domains is related to {\it{Markov Blanket}} discovery~\cite{Pearl88} in Bayesian networks and finds application in understanding latent structure of complex biological networks including heterogeneous bacterial mats such as those found in gut and lung flora~\cite{SuetalBIODATA-13}. There is also a growing literature on successfully using recurrent neural networks to solve related graph partitioning problems~\cite{PainetalIJNME-99,KechadiHPCN-99} also referred to as graph or spectral clustering~\cite{TianetalAAAI-14}.

We will likely start out by relying on a combination of backpropagation and some variant of AM as a solver. The loss function will be amended to include an additional term penalizing domain assignments in which two or more basis functions are assigned a nonzero assignment weight for the same point source. We may encounter technical difficulty with this part of the overall puzzle, but we have a reasonable strategy for starting out and have a identified a set of promising alternatives if our first attempts fail.





\label{sup_major_factors_important_technical_challenges}
\addcontentsline{toc}{section}{Supplement B: Principal Technical Challenges}
\section*{Supplement B: Principal Technical Challenges}


In this supplement we consider the major factors we believe will determine the success or failure of this endeavor. Roughly speaking they can be divided into challenges relating to the representation, implementation, execution, and evaluation or interpretation of the mesoscale models presented in this document. The representational challenges concern what sort of concepts and relationships can be described within the modeling language of the theory.

Implementation deals with how those concepts are realized in terms of data structures, algorithms and computations that perform inference. Execution refers to the feasibility (time and space) required for carrying out the computations necessary to infer mesoscale models and make predictions. Evaluation concerns how to determine whether the inferred models and their predictions are accurate and serve the intended purpose of providing a bridge between cellular and behavioral theories.

Insofar as representation can be reduced to the problem of learning a transfer function\footnote{%
 For the purposes of this discussion, a transfer function is a compact representation of the input-output relation of a dynamical system, where "compact" is the operative word, e.g., we will not accept and are not interested, even theoretically, in a table of all input-output pairs.},
the challenge concerns whether or not the class of mesoscale models proposed here can be used to model a neural microcircuit of sufficient complexity.

Given that an artificial neural network with at least one hidden layer, enough parameters and sufficient data and time for training is a universal function approximator\cite{CybenkoMCSS-89,HorniketalNEURAL-NETWORKS-89}, the theoretician's answer is unequivocally in the affirmative\footnote{%
  In the mathematical theory of artificial neural networks, the universal approximation theorem states that a feed-forward network with a single hidden layer containing a finite number of neurons\emdash{}a multilayer perceptron, can approximate continuous functions on compact subsets of $R^n$, under mild assumptions on the activation function. The theorem shows that simple neural networks can represent a wide variety of interesting functions when given appropriate parameters. ({\urlh{https://en.wikipedia.org/wiki/Universal_approximation_theorem}{SOURCE}})}.
The question gets more complicated when you ask whether the function is learnable in polynomial time\cite{ValiantCACM-84} or whether the system is observable\cite{Dorf1989} in the control-theoretic sense that the input provides accurate enough information about the underlying state of the microcircuit to account for what's going on inside the circuit\footnote{%
  In control theory, {\it{observability}} is a measure of how well the internal states of a system can be inferred from knowledge of its external outputs. The observability and controllability of a system are mathematical duals. ({\urlh{https://en.wikipedia.org/wiki/Observability}{SOURCE}})}.

We attempt to finesse the problem by recording not just inputs and outputs, but information relating to the activity of every {\it{neuron}} in the microcircuit. This approach is not entirely satisfactory as now we have to consider whether the recorded data, e.g., sparse estimates of local Ca$^{+2}$ concentrations, an imperfect proxy for what we really want, e.g., dense sampling of local field potentials, is sufficient to recover an accurate account of the underlying computations.

Here we appeal to the fact that our mesoscale modeling task does not require that we account for the microscale activity of every neuron. Rather we seek a mesoscale representation that assigns learned basis functions to selected sub networks to account for the aggregate behavior of ensembles of neurons spanned by such sub networks.

Representation also concerns the question of whether or not the model specified as a composition of functional modules serves to perform the central function of a mesoscale model, providing a coherent explanation that bridges the gap between the microscale interactions occurring within ensembles of individual neurons and macroscale observable behaviors taking place under reasonably accurate environmental conditions.

Explanatory value is largely in the eye (or mind) of the beholder and for some scientists artificial neural networks will always be inscrutable black boxes. It is our contention that shallow recurrent neural networks comprised of basic connectivity patterns / weight matrices and standard activation functions can be understood in terms of relatively straightforward computational primitives in much the same way that a sorting algorithm is comprehensible while its implementation, whether in terms of bit streams or sequences of machine instructions, is incomprehensible to all but the most committed.

Moreover, since we expect relatively few basis functions compared with the number of recorded neurons, we could in principle perform a postprocessing simplification step consisting of network distillation~\cite{HintonetalCoRR-15,RomeroetalCoRR-14} or, better yet, reconstitute each basis function as a phase-space reduction in a low-dimensional manifold of the original state space as is commonly done in dynamical systems theory. The latter approach is increasingly applied in computational neuroscience to obtain a compact summary of high-dimensional neural recordings in terms of phase-space reconstruction followed by dimensionality reduction and bifurcation structure~\cite{AljadeffetalNEURON-16,KatoetalCELL-15,RadulescuetalCMSB-15,SussilloandBarakNC-13}.

The challenges regarding computation refer to the availability of computing resources and the engineering time and (increasingly) energy expenditure required for training and testing models including the inevitable effort in parameter tuning. Given that learning a satisfactory (predictive, explanatory) mesoscale model will require training, testing and experimenting with a reasonably large number of not-so-satisfying models, as well as their subsequent use and evaluation in attempting to understand the particular target data set or organism, we need to consider full cost commensurate with some multiple of what's expected to infer a single model.

Our response to the computational challenges amounts to pointing out that we are already building models whose size in terms of the number of parameters and amount of training data rivals that of the models we propose here. We acknowledge that traditional measures of tractability based upon asymptotic worst-case runtimes are not the standard by which we should judge the anticipated computational effort. We routinely solve instances of computationally-complex problems that are classified NP hard if not worse. Conversely we routinely solve coomputationally-tractable problems classified as $O(n \log{} n)$ that are challenging simply because $n$ is so large.

Most of the algorithms that we use correspond to approximations and their runtimes are generally determined heuristically. Expert software engineers generally have a pretty good idea what will work to satisfy a given target of precision, recall and latency. Our experience at Google is that for problems of the type and size we are considering here involving target organisms with less than a million neurons and fewer than a billion synapses, we feel reasonably confident that good software engineering will prevail to find a way to generate solutions that are good enough to suit our purposes. That said, there are some unknowns that lead us to be cautious, if only to allocate sufficient time to finding tractable solutions.

The problem of learning mesoscale models requires our approximating some particularly difficult optimization problems that we mention here because we don't as yet have a clear idea about what sorts of trade-offs are tolerable with respect to what we need in order to learn satisfactory models. As an example, the problem of learning a sparse functional basis is generally characterized as a non-convex optimization problem, and algorithms for solving these problems employ approximations that are well enough characterized that scientists and engineers employ them routinely. 

Alternating minimization (AM) is a general method used for optimizing two variables jointly that is often applied to sparse coding problems. Given two variables $p$ and $q$, the method works by alternating between solving for $p$ and $q$; more precisely, keep $p$ constant and optimize for $q$, keep $q$ constant and optimize for $p$, and continue alternating until the algorithm converges. Expectation maximization (EM) is a special case of AM ~\cite{TusnadyandCsiszarSDSI-84}, and variants of AM are among the most efficient for solving this class of problems~\cite{AroraetalPMLR-15}.

In the case of sparse coding, $p$ might correspond to the weights of the linear combinations of basis functions used to reconstruct the data and $q$ to the configurable network parameters that instantiate the functional modules comprising the basis. If this were all there was to the problem, we could probably use some variant of AM to solve it~\cite{BaoetalPAMI-16,LeeetalNIPS-06}. However, in our case, the problem of learning a sparse basis is complicated by the fact that we also have to learn the functional module domains\footnote{%
  Throughout the document we've stretched the term "domain" to encompass both the domain and range of a function, its inputs and outputs. The elements of the domain and range are point sources corresponding to vertices in the connectome graph. Taken together, they define a minimal (cut) set of edges such that, if you remove those edges from the connectome graph, you obtain a disconnected sub graph containing the domain and range. In general, this sub graph contains not only the corresponding domain and range vertices, but all the remaining vertices reachable from the domain and range vertices / point sources in the isolated sub graph.}

Luckily we don't have to search the space of all possible domains, since we use a {\it{restriction bias}} to limit search. Specifically, we start with the set of all spherical volumes corresponding to a receptive field centered at a 3D point in the convolution grid. For each such volume, we consider the set of proposals corresponding the set of all possible completely enclosed sub graphs. The resulting set of all proposals for each point in the convolution grid is still large, but significantly smaller than the set of all possible sub graphs of the full connectome graph. 

The resulting {\it{domain assignment problem}} is at least as hard as the {\it{graph partition}} problem in theoretical computer science which is known to be NP-hard\cite{FederetalSTOC-99,GareyandJohnson79}. That is not surprising, nor is it particularly discouraging. We are assuming that there is a certain amount of redundancy in the neural microcircuit and hence there will likely be multiple solutions and we are not committed to finding the optimal solution. Happily there is a growing literature on successfully using recurrent neural networks to solve related graph partitioning problems~\cite{PainetalIJNME-99,KechadiHPCN-99} also referred to as graph or spectral clustering in the literature~\cite{TianetalAAAI-14}.

We will likely start by relying on a combination of backpropagation and some variant of AM as a solver. The loss function will be amended to include an additional term penalizing domain assignments in which two or more basis functions are assigned a nonzero assignment weight for the same point source, e.g., $\hmlambda{}_{R} \hmtimes{} {\mbox{\rm{reconstruction error}}} + \hmlambda_{B} \hmtimes{} \hmsum{} {\mbox{\rm{basis combination weights}}} +  \hmlambda_{D} \hmtimes{} \hmsum{} {\mbox{\rm{domain assignment weights}}}$. We may encounter technical difficulty with this part of the overall puzzle, but we have a reasonable strategy for starting out.

Finally, we consider a few of the challenges relating the evaluation, validation and interpretation of mesoscale models. We have an aggressively data-driven machine learning (ML) approach to scalable neuroscience. Insofar as possible, we avoid the need for human interpretation and hand annotation of data, and regarding the challenges relating to model validation we adopt a combination of traditional ML testing on held out data and expert analysis conducted in collaboration with our partner institutions.

Since we rely heavily on data, a word about what sort of data we require and when is in order. We need both structural and functional data. The former involves the use of electron microscopy to reconstruct all of the neuropil in a tissue sample. Thanks to state-of-the-art hardware and a good deal of effort, we are reasonably confident we will be able to obtain the structural data we need when we need it. Functional recording depends on being able to record from all or most of the neurons in the brains of awake behaving animals, and while the relevant technology is not as mature as we find in electron microscopy, the following projections bode well for acquiring the necessary functional data in a timely manner:
\begin{itemize}
\item Time frame 0-2 years: Dense, pan-neuronal, $\approx{}100,00 neurons$, single-cell resolution, using (a) light-field imaging, $\approx{}30$ FPS; (b) multi-photon excitation microscopy and light-sheet scanning, $\approx{}$5 FPS; (c) image post-processing software, including sparse and unique nonnegative matrix factorization, and superresolution strategies including PALM (photoactivated localization microscopy) and STORM (stochastic optical reconstruction microscopy).
\item Time frame 2-5 years: Improved resolution, new reporters targeting synapses, genetically encoded voltage indicators\emdash{}earlier work depends heavily on calcium reporters in the cell body often the nucleus, imaging multiple areas simultaneously, e.g., V1, V2, IT in mammal ventral visual pathway, using multiple piezoelectric-slaved excitation beams, improvements in superresolution structured illumination microscopy (SR-SIM) leveraging very high-speed cameras, new techniques for working with freely moving organisms\footnote{%
    Microlens arrays, very high-frame rate\emdash{}more than a trillion FPS\emdash{}camera sensors, and other technical improvements have yet to be incorporated into microscopes used for functional imaging. Word spreads fast and equipment manufacturers are realizing they can charge more than \$1M dollars for bleeding edge functional recording technology. We are cautiously optimistic that the industry will meet or exceed our expectations for improved functional and structural data acquisition.}.
\end{itemize}

There is one crucial dimension of mesoscale modeling in which more data and better machine learning technology will not carry the day, and that concerns the requirement that mesoscale models serve as explanatory theories. They are intended first and foremost as tools for asking questions, exploring hypotheses, gaining insight and ultimately inspiring new hypotheses and research. The sort of tools we are developing will be judged on the basis of how well they enable scientists to do good science.

Successful partnering means technologist tool builders and scientist domain experts work together to shape the language in which mesoscale models are expressed, test existing hypotheses and formulate new ones, figure out how best to use and improve the technology for collecting and annotating data at scale, and rapidly accommodate new technologies that enable new behavioral, e.g., Robie~\etal{}~\cite{RobieetalCELL-17}, anatomical, e.g., Seiriki~\etal{}~\cite{SeirikietalNEURON-17}, and functional, e.g., Nobauer~\etal{}~\cite{NobaueretalNATURE-METHODS-17} and Kim~\etal{}~\cite{KimetalNATURE-METHODS-17} insights.



\end{document}